\documentclass[twocolumn ,aps,prd,a4paper,superscriptaddress,tightenlines,noeprint,10pt,floatfix,nobibnotes]{revtex4-1}
\RequirePackage{float}
\usepackage[normalem]{ulem}
\usepackage{subcaption}
\usepackage[utf8]{inputenc}
\usepackage[USenglish]{babel}
\usepackage[T1]{fontenc}
\usepackage{graphicx,epsf}
\usepackage[toc,page,header]{appendix}
\usepackage{minitoc}
\usepackage{tabularx}
\usepackage{color}
\usepackage[bookmarksnumbered,hypertexnames=false,colorlinks,colorlinks=true,linkcolor=blue,urlcolor=black,citecolor=blue,anchorcolor=green]{hyperref}
\usepackage{bbm,braket,microtype,mathrsfs,amsmath,amssymb,color,amsthm,mathtools,fullpage,enumitem,ae,aecompl,bm,thm-restate}
\usepackage{physics}
\usepackage{tikz}
\usepackage{lipsum}
\usepackage{nameref}
\usepackage{color}
\usepackage{booktabs, caption, palatino}
\usepackage[capitalize]{cleveref}
\usepackage{silence}

\usepackage{mathtools}

\allowdisplaybreaks
\newcommand{\be}{\begin{equation}}
\newcommand{\ee}{\end{equation}}
\newcommand{\ba}{\begin{eqnarray}}
\newcommand{\ea}{\end{eqnarray}}

\newcommand\stab{{\operatorname{STAB}}}
\newcommand{\ignore}[1]{}

\newcommand{\ot}{\otimes}

\newcommand{\pauli}[1]{\mathbb{P}_{#1}}

\newcommand{\absval}[1]{\left| {#1} \right|}

\newcommand{\Id}{\mathbbm{1}}

\def\CC{{\rm\kern.24em \vrule width.04em height1.46ex depth-.07ex
   \kern-.29em C}}
\def\P{{\rm I\kern-.25em P}}
\def\RR{{\rm
        \vrule width.04em height1.58ex depth-.0ex
        \kern-.04em R}}

\def\bbbc{{\mathchoice {\setbox0=\hbox{$\displaystyle\rm C$}\hbox{\hbox
to0pt{\kern0.4\wd0\vrule height0.9\ht0\hss}\box0}}
{\setbox0=\hbox{$\textstyle\rm C$}\hbox{\hbox
to0pt{\kern0.4\wd0\vrule height0.9\ht0\hss}\box0}}
{\setbox0=\hbox{$\scriptstyle\rm C$}\hbox{\hbox
to0pt{\kern0.4\wd0\vrule height0.9\ht0\hss}\box0}}
{\setbox0=\hbox{$\scriptscriptstyle\rm C$}\hbox{\hbox
to0pt{\kern0.4\wd0\vrule height0.9\ht0\hss}\box0}}}}
\def\bbbz{{\mathchoice {\hbox{$\sf\textstyle Z\kern-0.4em Z$}}
{\hbox{$\sf\textstyle Z\kern-0.4em Z$}}
{\hbox{$\sf\scriptstyle Z\kern-0.3em Z$}}
{\hbox{$\sf\scriptscriptstyle Z\kern-0.2em Z$}}}}

\newlength{\fighskip} \fighskip=2pt
\newlength{\figvskip} \figvskip=1pt

\makeatletter
\def\namedlabel#1#2{\begingroup
   \def\@currentlabel{#2}%
   \label{#1}\endgroup
}
\makeatother
\usepackage{hyperref}

\begin{document}
\setcounter{secnumdepth}{3}
\title{Harvesting stabilizer entropy and non-locality from a quantum field}
\author{Simone Cepollaro}
\email[Corresponding author: ]{simone.cepollaro-ssm@unina.it}
\affiliation{Scuola Superiore Meridionale, , Largo S. Marcellino 10, 80138 Napoli, Italy}
\affiliation{INFN, Sezione di Napoli, Italy}
\author{Stefano Cusumano}
\affiliation{INFN, Sezione di Napoli, Italy}
\affiliation{Dipartimento di Fisica `Ettore Pancini', Universit\`a degli Studi di Napoli Federico II,
Via Cintia 80126,  Napoli, Italy}

\author{Alioscia Hamma}
\affiliation{Scuola Superiore Meridionale, , Largo S. Marcellino 10, 80138 Napoli, Italy}
\affiliation{INFN, Sezione di Napoli, Italy}
\affiliation{Dipartimento di Fisica `Ettore Pancini', Universit\`a degli Studi di Napoli Federico II, Via Cintia 80126,  Napoli, Italy}

 \author{Giorgio Lo Giudice }
\affiliation{Dipartimento di Fisica `Ettore Pancini', Universit\`a degli Studi di Napoli Federico II,
Via Cintia 80126,  Napoli, Italy}

\author{Jovan Odavi\' c}
\affiliation{INFN, Sezione di Napoli, Italy}
\affiliation{Dipartimento di Fisica `Ettore Pancini', Universit\`a degli Studi di Napoli Federico II,
Via Cintia 80126,  Napoli, Italy}

\begin{abstract}
The harvesting of quantum resources from the vacuum state of a quantum field is a central topic in relativistic quantum information. While several proposals for the harvesting of entanglement from the quantum vacuum exist, less attention has been paid to other quantum resources, such as non-stabilizerness, commonly dubbed {\em magic} and quantified by the Stabilizer R\'enyi Entropy (SRE). In this work, we show how to harvest SRE from the vacuum state of a massless field\sout{,} using accelerated Unruh-DeWitt detectors in Minkowski spacetime. In particular, one can harvest a particular non-local form of SRE that cannot be erased by local unitary operations. This non-local SRE is a fundamental quantity to study the interplay between entanglement and non-stabilizer resources. We conclude our work with an analysis of the CHSH inequalities: when restricting to stabilizer measurements, i.e. Pauli measurements, one cannot extract a violation from the quantum field.
\end{abstract}

\maketitle

\section{Introduction}

The harvesting of quantum resources from the vacuum state of a quantum field~\cite{PhysRevD.92.064042,PhysRevD.108.085025,PhysRevD.78.045006,Stritzelberger_2021} plays a prominent role in the interplay between quantum information and quantum field theory, due to its applications in quantum information processing~\cite{PhysRevD.101.036014,kasprzak2024transmissionquantuminformationquantum}, quantum gravity~\cite{PhysRevLett.119.240401,PhysRevLett.119.240402,Huggett_Linnemann_Schneider_2023,PhysRevD.108.L101702,PhysRevD.101.125018,Mart_n_Mart_nez_2014,PhysRevD.111.025015}, black holes physics~\cite{wang2025harvestinginformationhorizon,PhysRevD.106.025010}, and newly discovered phenomena such as energy teleportation~\cite{doi:10.1143/JPSJ.78.034001,PhysRevD.81.044025}.

Harvesting consists in extracting or unlocking a quantum resource originally located in a quantum system, e.g., a quantum field, where it cannot be exploited, and moving it into another quantum system, e.g., a detector, from which the harvested resource can then be used to perform useful tasks. The most prominent and known example of these protocols is entanglement harvesting: the vacuum state of a free field is entangled with respect to the bipartition given by two accelerating observers~\cite{Unruh_1976,VALENTINI1991321,Martín-Martínez_2014} and through an entanglement harvesting protocol it is possible to move the entanglement from the field to a pair of detectors.
This kind of protocol has been analyzed in many different settings~\cite{PhysRevD.102.125026,PhysRevD.103.016007,PhysRevD.106.076002}, as one can change the kind of detector used for harvesting~\cite{LSriramkumar_1996,DeSLTorres2023} or consider causally disconnected~\cite{Reznik_2003} or accelerated detectors~\cite{Pozas_2015,Salton_2015}.

Beyond entanglement, more general quantum resources can be harvested from quantum fields~\cite{PhysRevA.88.062336,PhysRevD.107.065016,PhysRevD.108.105017,Lin2024,nystrom_2024_harvestingmagic}. In this work, we propose a novel protocol aiming at harvesting non-stabilizerness, colloquially dubbed {\it magic}. Non-stabilizerness is the fundamental resource necessary to achieve quantum advantage in computational tasks, and it has thus received great attention
for its role in quantum information processing and quantum complexity \cite{10.21468/SciPostPhys.9.6.087,Leone2021quantumchaosis,catalano2025magicphasetransitionnonlocal,jasser2025stabilizerentropyentanglementcomplexity}.
Indeed, from a complexity perspective, even maximally entangled states can be efficiently simulated using classical resources~\cite{gottesman_theory_1998,gottesman_demonstrating_1999,aaronson_improved_2004}, as long as they do not possess resources beyond stabilizer ones.  

As non-stabilizerness is a fundamental resource to achieve quantum advantage, many measures for its quantification have been proposed~\cite{PhysRevLett.118.090501,PhysRevLett.124.090505,Regula_2018,Beverland_2020}, such as the Wigner negativity~\cite{Veitch_2014}, the stabilizer rank~\cite{Bravyi2019simulationofquantum} and the Stabilizer R\'enyi Entropy (SRE)~\cite{Leone_2022_SRE}. In particular, the SRE is the unique computable magic monotone for pure states \cite{Leone_2024} and it is computed from the expectation values of a quantum state over the Pauli strings. SRE can be computed in an efficient way by perfect sampling \cite{PhysRevLett.131.180401, Haug_2023} and can be extended to both qudits \cite{Wang2023} and mixed states. This resource is also experimentally measurable \cite{Oliviero_2022}. 

Non-stabilizerness is a central quantity in scenarios that go beyond quantum information processing. It has been proved that it plays a role in  Conformal Field Theories ~\cite{PhysRevB.103.075145} and in processes of state decoding from a black hole ~\cite{PatrickHayden_2007,PhysRevA.106.062434}. Non-stabilizerness has been recently shown to be a necessary ingredient in the context of simulations of quantum gravity models~\cite{cao2024gravitationalbackreactionmagical,PhysRevD.109.126008}, nuclear physics \cite{Brokemeier:2024lhq,Robin:2024bdz} and dense neutrinos systems \cite{Chernyshev:2024pqy}. 

If one restricts to Pauli measurements, non-stabilizer resources are also necessary to violate the Clauser-Horne-Shimony-Holt (CHSH) inequalities \cite{PhysRevA.91.042103}. This restriction corresponds, in a resource theoretic spirit, to the restriction to local measurements in establishing the importance of entanglement for such violations~\cite{cusumano2025nonstabilizernessviolationschshinequalities}. 

In this paper, we study the harvesting of both SRE and entanglement  from the vacuum of a scalar field. We start our analysis from an inertial setting, where some analytic expressions can be derived, before turning our attention to an accelerated reference frame. We show that both entanglement and SRE can be harvested from the scalar field. In particular, the harvested SRE comes in a completely non-local form called non-local stabilizer entropy $M^{NL}$. This quantity is non-local in the sense that it cannot be erased by local unitary operations \cite{cao2024gravitationalbackreactionmagical}. This is a particular form of SRE that can only exist in entangled states, but not in all of them, notably never in maximally entangled states.

At this point, it comes natural to ask whether the harvested non-local resources can result indeed in non-local effects as a violation of the  CHSH inequalities \cite{Summers1985,Summers1987,guimaraes2025bellsinequalityrelativisticquantum,PhysRevD.110.105001}. The answer is  that there is no way that one, starting with a resource-free state can violate CHSH inequalities by resource harvesting. Moreover, even if one starts with an initial state containing one of the two resources, there is no way of extracting
a violation of CHSH inequalities. In \cite{TIAN201298,wu_does_2024,Guedes_2024}, the authors study the behavior of the Bell operator for two detectors interacting with quantum fields, but not with the goal of harvesting. Starting with a state that violates CHSH inequalities, they find that the violation decreases for the interaction with a quantum field. The present  paper will also explain why a certain kind of coupling between field and detectors will not allow for non-locality detection.

In~\cite{nyström2024harvestingmagicvacuum} a magic harvesting protocol has been proposed using a three-level Unruh-DeWitt detector and a quantifier known as mana~\cite{Veitch_2014}. Our work differs from~\cite{nyström2024harvestingmagicvacuum} and goes beyond it in at least two aspects. 

First, as clarified by the authors in~\cite{nyström2024harvestingmagicvacuum} the appearance of non-stabilizerness in the detector's final state is due {\em solely} to the interaction of the detector with the quantum field, so there is no actual harvesting of the resource. In this paper, we harvest the non-stabilizer resource from the field in two ways: in a {\em weak sense},  because we show that the protocol would not be able to induce non-stabilizer resources if the quantum field did not possess them and, in a {\em strong sense}, that the extraction of a particularly strong form of SRE, the so-called {\em non-local} SRE\cite{cao2024gravitationalbackreactionmagical} is extracted through a protocol that is a free operation for the associated resource theory. These notions are cognate to that of resource {\em embezzlement}, see \cite{vanluijk2024embezzlingentanglementquantumfields}.

 Second, as we use SRE in place of  mana to quantify non-stabilizerness, we are able to consider arbitrary states of the detectors. This allows us to consider a different physical setting, e.g. two causally disconnected observers moving with two accelerating reference frames with parallel, antiparallel and perpendicular mutual acceleration.
\begin{figure*}[!t]
\centering
\begin{subfigure}{0.33\textwidth}
\centering
\includegraphics[width=\linewidth]{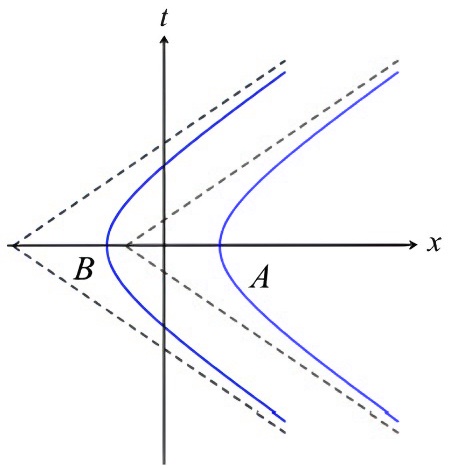}
\caption{}
\label{fig:par_wl}
\end{subfigure}
\begin{subfigure}{0.33\textwidth}
\centering
\includegraphics[width=\linewidth]{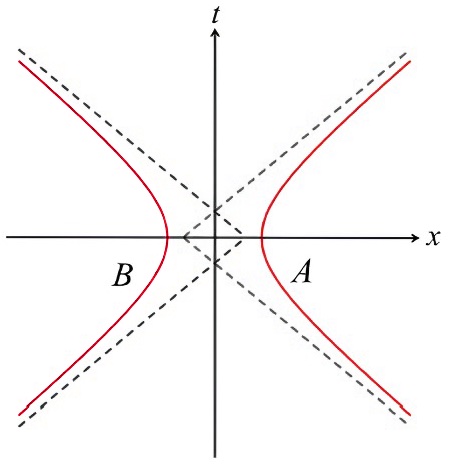}
\caption{}
\label{fig:antipar_wl}
\end{subfigure}
\begin{subfigure}{0.33\textwidth}
\centering
\includegraphics[width=\linewidth]{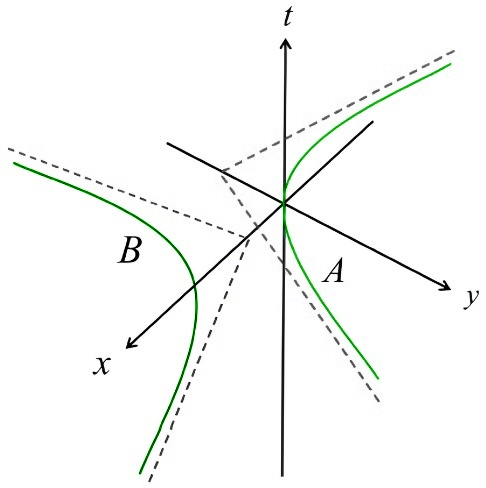}
\caption{}
\label{fig:perpendicular_wl}
\end{subfigure}
\caption{Trajectories of accelerated UDW detectors: (a) detectors with parallel acceleration $a_{\parallel}$; (b) detectors with antiparallel acceleration $a_{\nparallel}$; (c) detectors with perpendicular acceleration $a_{\perp}$.}
\label{fig:worldlines}
\end{figure*}

\section{Setting}
\subsection{The harvesting protocol}

We consider two point-like detectors $A$ and $B$ moving in an accelerated reference frame with respect to each other. These are modeled as two level systems with Hamiltonian:
\ba
\nonumber
H_{AB}=H_A+H_B=\frac{\Omega}{2}Z_A+\frac{\Omega}{2}Z_B,
\ea
where $Z_i$ with $i=A,B$ is the diagonal Pauli operator acting on the Hilbert space of system $A$ and $B$ respectively.
We consider a massless scalar field $\phi$ in Minkowski space-time described by the free Hamiltonian:
\ba
\nonumber
H_\phi=\int \frac{d^3\bm{p}}{(2\pi)^3}\absval{\bm p} a^\dag_{\bm p} a_{\bm p}.
\ea
where $a^\dag_{\bm p}$ and $a_{\bm p}$ are the creation and annihilation operators obeying canonical commutation relations, and ${\bm p}$ is the momentum.

The two detectors interact with the quantum field via the interaction Hamiltonian (in the interaction picture):
\ba
\label{eq:int_hamiltonian}
H_{\rm int}^{(i)}=\lambda\epsilon_i(\tau)\mu_i(\tau)\otimes\phi(\bm{x}_i(\tau)),
\ea
where $i=A,B$ and $\tau$ is the proper time. Here $\lambda<<1$ is a parameter gauging the strength of the interaction and $\epsilon_i(\tau)$ is a switching function describing the turning on and off of the interaction. The operator $\phi(\bm{x}_i(\tau))$ is the usual field operator whose Fourier expansion is:
\ba 
\nonumber
\phi(\bm{x})=\int \frac{d^3 \bm{p}}{(2\pi)^3} \frac{1}{\sqrt{2\absval{\bm{p}}}}\left(a_{\bm p}e^{-i\bm{p} \cdot \bm{x}} + a^{\dag}_{\bm p}e^{i\bm{p}\cdot \bm{x}}\right)
\ea 
evaluated along the space-time trajectory of detector $i$. Finally, the operator $\mu_i(\tau)$ acts on the detector as:
\ba
\nonumber
\mu_i(\tau)&=&e^{+i\Omega\tau}\dyad{1}{0}+e^{-i\Omega\tau}\dyad{0}{1}\\
\nonumber
&=&e^{+i\Omega\tau}\sigma_i^{+}+e^{-i\Omega\tau}\sigma_i^{-}.
\ea

Notice that in contrast with entanglement harvesting protocols, in principle one does not need two detectors to harvest non stabilizer resources, since this resource can be highly non-trivial even for a single qubit. Here, we still use two detectors. This choice allows us to study also the interplay between entanglement and non-stabilizerness as revealed in the non-local SRE and in their relationship with violations of locality in the setting of the CHSH inequalities.

\subsection{Entanglement and SRE}
To quantify the entanglement of the state $\rho_{AB}$ of the detectors we use  the concurrence $\mathcal{C}$~\cite{PhysRevLett.78.5022,PhysRevLett.80.2245}, which for the case of two qubits reads:
\ba
\nonumber
\mathcal{C} :=\max\{0,\lambda_1-\lambda_2-\lambda_3-\lambda_4\}
\ea
where $\lambda_i$ are the eigenvalues, in decreasing order, of the matrix $R=\rho\tilde{\rho}$ and
$\tilde{\rho}=(Y\otimes Y)\rho^* (Y\otimes Y)$. Let us also notice that for the case of two qubits any measure of entanglement is equivalent, so that using other measures of entanglement, such as entanglement entropy or logarithmic negativity would lead to the same results~\cite{RevModPhys.81.865}. As the use of entanglement measures in harvesting protocols has been widely described in the literature on entanglement harvesting, we refer the reader the the references already cited in the introduction for further details on entanglement measures and harvesting.

Let us now focus on the other quantity of interest for our analysis, the Stabilizer R\'enyi Entropy ~\cite{Leone_2022_SRE}.
Loosely speaking the SRE is a measure of how much the state of a quantum system is ``spread'' over the Pauli basis. For pure states, it is the {\em unique} computable monotone \cite{Leone_2024} for non-Clifford resources that go beyond the stabilizer formalism. Both quantum advantage and quantum complexity require these resources. Indeed, stabilizer states and Clifford operations play a central role in the framework of quantum computation: the Gottesman-Knill theorem \cite{gottesman_theory_1998} states that any quantum process involving only stabilizer states and Clifford operations can be efficiently simulated classically in polynomial time. For this reason, they would also not be able to offer any quantum advantage.

Define now $\mathbb{P}_n$ the set of all n-qubit Pauli strings. A Pauli string is an n-fold tensor product of Pauli operators, that is $\mathbb{P}_n=\mathbb{P}_1^{\otimes n}$ with $\mathbb{P}_1=\{\Id,X,Y,Z\}$. The n-qubit Pauli group $\mathcal{P}_n$ is then defined as the group formed by the elements of $\mathbb{P}_n$ multiplied by the phases $\{\pm1,\pm i\}$.
The \textit{Clifford group} $\mathcal{C}_n$ is the normalizer of the Pauli group, that is the set of all n-qubit unitary operators $C$ mapping an element of the Pauli group into an element of the Pauli group, or more formally:
\ba
\nonumber
\mathcal{C}_n=\{ C\in\mathcal{U}(n):\forall P\in \mathcal{P}_{n},CPC^{\dag}=P'\in \mathcal{P}_{n}\}.
\ea 
The set of pure \textit{stabilizer states} is defined as the Clifford orbit of any of the computational basis states $\{\ket{i}\}$, namely
\ba
\nonumber
\stab=\{C\ket{i},C\in\mathcal{C}(n)\}.
\ea
Notice that the Clifford group is a discrete subgroup of the Unitary group, so that for any number of qubit $n$ the cardinality $|\rm{STAB}|$ of the set of stabilizer states is finite. 

The $\alpha$-SRE $M_\alpha(\ket{\psi})$ of an n qubits pure state $\ket{\psi}$ is defined as the $\alpha$ Rényi entropy of the probability distribution defined by \cite{Leone_2022_SRE}:
\ba
\nonumber
\Xi_P(\ket{\psi})=\frac{\bra{\psi}P\ket{\psi}^2}{d},\;P\in\mathbb{P}_n,
\ea
where $d=2^n$ is the dimension of the Hilbert space, so that the $\alpha$-SRE can be written as:
\ba
\nonumber
M_{\alpha}(\ket{\psi})=(1-\alpha)^{-1}\log\sum_{P\in\mathbb{P}_n}\Xi_P^\alpha(\ket{\psi})-\log d
\ea
In this work we use the $2$-SRE $M_2$ to quantify the amount of non-stabilizerness in the detectors final state , which for pure states turns out to be:
\ba
\nonumber
M_2(\ket{\psi})&=& -\log\left[d\sum_{P\in \mathbb{P}_n} \Xi_P(\psi)^2\right]\\
\nonumber
&=&-\log \left( d^{-1} \Tr^4 (\psi P) \right)\\
&=&-\log d\Tr{Q\dyad{\psi}^{\otimes4}}
\ea
where $Q:=d^{-2}\sum_{P\in\pauli{n}}P^{\otimes4}$.

The 2-SRE can be extended to  mixed states:
\ba
\label{eq:sre_mixed}
\tilde{M}_2(\rho) :=M_2(\rho)-S_2(\rho),
\ea
where $M_2(\rho)=-\log d\Tr{Q\rho^{\ot4}}$ and $S_2(\rho)=-\log\Tr{\rho^2}$ is the 2-R\'enyi entropy of $\rho$.
Plugging the expression for the final state of the detectors $\rho_{AB}$ into Eq.~\eqref{eq:sre_mixed} one can find an expression for the SRE $\tilde{M}_2(\rho_{AB})$  as a function of the acceleration $a$. Further details are reported in Sec.~\ref{sec:perturbative_SRE}. 

It is important to remark that for mixed states  SRE does not have the usual meaning of distillable non-stabilizer resources. In this sense, it behaves exactly like the von Neumann entropy for entanglement. SRE has always, though, the operational meaning of quantifying the hardness of certain quantum information protocols \cite{PhysRevA.107.022429, PhysRevLett.132.080402, PhysRevA.109.022429}. Moreover, the resource free states $\tilde{M}_2(\rho)=0$ are those that cannot be purified in a stabilizer state\cite{cao2024gravitationalbackreactionmagical, gottesman_theory_1998}. The quantity $\tilde{M}_2$ is in fact a good {\em proxy} for this resource theory\cite{prep}.

The {\em non-local} stabilizer entropy can be defined \cite{cao2024gravitationalbackreactionmagical} as
\ba
\nonumber
M_\alpha^{NL} (\ket{\psi}):= \min_{R=U_A\otimes U_B} M_\alpha(R \ket{\psi})\\
\nonumber
M_\alpha^{NL} (\rho):= \min_{R=U_A\otimes U_B} M_\alpha(R \rho R^\dag)
\ea
for any $\alpha$-SRE and for both pure and mixed states. $M_\alpha^{NL}$ represents the non-stabilizer resources that cannot be erased by local unitary operations. It plays an important role in the AdS-CFT correspondence, as it represents the holographic dual of gravitational back-reaction \cite{cao2024gravitationalbackreactionmagical}. As $M_\alpha(\psi)^{NL}$ vanishes identically on both product and stabilizer states, it quantifies the non-locality of SRE and serves as a useful probe to investigate the interplay between entanglement and non-stabilizer resources. Notice that a LOCC protocol would not be able to extract non-local SRE $M_\alpha^{NL}$. Thus, to the extent that there is genuine harvesting of entanglement, there is genuine harvesting of $M^{NL}$ in the strong sense. While SRE could in principle be extracted by the field even by a single detector, and this would not say anything about the non-locality of the field, extracting non-local SRE necessitates two detectors interacting (locally) with an entangled field. 

\subsection{CHSH inequalities}
In this section, we analyze the role played by 
non-stabilizerness  in order to violate the CHSH formulation of Bell's inequality \cite{howard_nature, PhysRevA.91.042103}. Starting with the state $\omega=\dyad{00}$ and considering a Bell operator $\mathcal{B}_0$ of the form
\ba
\nonumber
\mathcal{B}_0=P_A^{(1)}\left(P_B^{(1)}+P_B^{(2)}\right)+P_A^{(2)}\left(P_B^{(1)}-P_B^{(2)}\right)
\ea
with all the $P_i^{(j)}\in\{X,Y,Z\}$, one easily obtains:
\ba
\label{eq:bell_bound}
b_0=\rm{Tr}\left[\mathcal{B}_0C\omega C^\dag\right]\leq2
\ea
where $C$ is any unitary operation belonging to the Clifford group. The meaning of Eq.~\eqref{eq:bell_bound} is that entanglement alone is not sufficient to violate the CHSH inequality. Notice that this is not in contrast with the usual setting of Bell's inequality in the CHSH setting, where measurements in any direction are allowed, which in general require non stabilizer resources   to be performed~\cite{cusumano2025nonstabilizernessviolationschshinequalities}.
One is then tempted to check whether it is possible to harvest enough SRE and entanglement as to violate the CHSH inequality, since this would imply the possibility of harvesting non-locality.

In the following, we  restrict our measurements to be both  local and Pauli. In this way, there are no resources that go beyond Clifford or locality. We thus define a Bell operator 
\be 
\label{eq:bell_operator}
\mathcal{B}_0 :=X_AX_B+X_AZ_B-Z_AX_B+Z_AZ_B
\ee 
whose maximum value  over the set of stabilizer states $\tr (\mathcal{B}_0\Phi^+)=2$ is obtained for the maximally entangled state $\ket{\Phi^+}=(\ket{00}+\ket{11})/\sqrt{2}$ and it falls just short of violating the CHSH inequalities.  We indicate with $b_0\equiv \Tr[\rho_{AB}(0)\mathcal{B}_0]$ the value of the Bell operator at the initial time $t=0$.

\section{Perturbative expansion and technical results}
In this section we are going to derive the explicit expression for the CPT map mapping the initial state of the detectors into the final state after the interaction with the field has taken place. Furthermore, we are going to derive the explicit perturbative expressions for the concurrence and the SRE, up to order $\mathcal{O}(\lambda^2)$. We will make explicit example for the most significant case of the two detectors starting in the state $\ket{00}$, but analogous formulas can be derived for any initial state of the field.
\subsection{CPTP map of the detector}

The state of the joint system evolves in the interaction picture according to the Hamiltonian in Eq.~\eqref{eq:int_hamiltonian}, so that we can write the unitary operator $U$ describing the dynamics of the system as:
\ba
\label{eq:unitary_operator}
U=\mathcal{T}\exp\left[-i\left(\int d\tau_AH_{\rm int}^{(A)}+\int d\tau_BH_{\rm int}^{(B)}\right)\right].
\ea
One can then compute the final state of the joint system as $\ket{\psi_f}=U\ket{\psi_0}$. Since we are interested in the resources of the detector state, we can  trace out the field degrees of freedom and obtain the final (mixed) state of the detectors
\ba
\label{eq:CPT_definition}
\rho_{AB}=\Tr_{\phi}\left[U\rho(0)U^{\dag}\right]:=\mathcal{E}(\rho_{AB}(0))
\ea
where $\rho(0)=\rho_{AB}(0)\otimes\dyad{0_\phi}$.

In order to derive the Completely Positive Trace Preserving (CPTP) superoperator $\mathcal{E}$  mapping the initial state of the detectors into the final one, we first expand perturbatively, up to the order $\lambda^2$, the operator $U$, obtaining:
\begin{widetext}
\ba
    {U} &=& {\mathcal{T}} \exp\left[-i   \int dt \Bigl(\frac{d\tau_A}{dt} {H}_{\rm int}^{(A)} (\tau_{A}(t)) + \frac{d\tau_B}{dt} {H}_{\rm int}^{(B)} (\tau_{B}(t))\Bigr)\right] \nonumber \\
    &=& {{\mathbbm{1}}}-i \int dt \Bigl(\frac{d\tau_A}{dt} {H}_{\rm int}^{(A)} (\tau_A(t)) + \frac{d\tau_B}{dt} {H}_{\rm int}^{(B)} (\tau_B(t))\Bigr) \nonumber\\
    &-&\frac{1}{2} \int dt \int dt' {\mathcal{T}} \Bigl[ \Bigr. \frac{d\tau_A}{dt} {H}_{\rm int}^{(A)}(\tau_{A}(t)) \frac{d\tau_A}{dt'} {H}_{\rm int}^{(A)}(\tau_{A}(t'))+
\frac{d\tau_B}{dt} {H}_{\rm int}^{(B)}(\tau_{B}(t)) \frac{d\tau_B}{dt'} {H}_{\rm int}^{(B)}(\tau_{B}(t')) \nonumber \\
\label{eq:U_power_expansion}
&+& \frac{d\tau_A}{dt} {H}_{\rm int}^{(A)}(\tau_{A}(t)) \frac{d\tau_B}{dt'} {H}_{\rm int}^{(B)}(\tau_{B}(t'))+ \frac{d\tau_B}{dt} {H}_{\rm int}^{(B)}(\tau_{B}(t)) \frac{d\tau_A}{dt'} {H}_{\rm int}^{(A)}(\tau_{A}(t')) \Bigl. \Bigr] 
\ea
\end{widetext}
Notice that we have written $U$ in terms of the coordinate time $t$, with respect to which the vacuum state of the field is defined.
What is left to do is to insert the power expansion of $U$ from Eq.~\eqref{eq:U_power_expansion} into Eq.~\eqref{eq:CPT_definition}.
To simplify the notation, we write the operator $U$ as:
\ba
\nonumber
U=U_0+U_1+U_2+\mathcal{O}(\lambda^3)
\ea
where $U_i$ are just the corresponding powers of $\lambda$ in Eq.~\eqref{eq:U_power_expansion}.
Tracing away the field degrees of freedom one obtains:
\ba
\nonumber
\mathcal{E}(\rho_{AB})&=&\Tr_{\phi}\left[U_0\rho(0)U_0^\dag\right]\\
\nonumber
&+&\Tr_{\phi}\left[U_1\rho(0)+\rho(0)U_1^\dag\right]\\
\nonumber
&+&\Tr_{\phi}\left[U_1\rho(0)U_1^\dag+U_2\rho(0)+\rho(0)U_2^\dag\right]
\ea
Let us then compute all the terms, starting from the zero-th order one:
\ba
\nonumber
\Tr_{\phi}\left[U_0\rho(0)U_0^\dag\right]=\rho_{AB}(0).
\ea
The first order can be derived by considering the term $U_1\rho(0)$:
\ba
\nonumber
&&U_1\rho(0)=\\
\nonumber
&&-i \int dt \Bigl(\frac{d\tau_A}{dt} {H}_{\rm int}^{(A)} (\tau_A(t)) + \frac{d\tau_B}{dt} {H}_{\rm int}^{(B)} (\tau_B(t))\Bigr)\rho(0)
\ea
This term, similarly to its Hermitian conjugate, will be null after tracing away the field, since in the expression above all the terms have only one field operator and the state of the field is diagonal in the Fock basis.

We are thus left with the second order contribution. Let us start from the term $U_1\rho(0)U_1^\dag$:
\begin{widetext}
\ba
\nonumber
U_1\rho(0)U_1^\dag=\left[\int dt \Bigl(\frac{d\tau_A}{dt} {H}_{\rm int}^{(A)} (\tau_A(t)) + \frac{d\tau_B}{dt} {H}_{\rm int}^{(B)} (\tau_B(t))\Bigr)\right]\rho(0)\left[\int dt' \Bigl(\frac{d\tau_A}{dt'} {H}_{\rm int}^{(A)} (\tau_A(t')) + \frac{d\tau_B}{dt'} {H}_{\rm int}^{(B)} (\tau_B(t'))\Bigr)\right]\\\label{eq:first_order_U11}
\ea
\end{widetext}
This time bilinear combinations of field operators will appear, so that some of the terms will not be null after tracing away the field.
To derive a more explicit expression, let us study explicitly the perturbative terms appearing in Eq.~\eqref{eq:first_order_U11}. Specifically, let us study the term:
\ba
\nonumber
\left(\int dt \frac{d\tau_A}{dt} {H}_{\rm int}^{(A)} (\tau_A(t))\right)\rho(0)\left(\int dt' \frac{d\tau_A}{dt'} {H}_{\rm int}^{(A)} (\tau_A(t'))\right)
\ea
as all the other combinations of $H_{\rm int}^{(i)}$ can be worked out in the same way.
Writing explicitly the interaction Hamiltonian we obtain:
\begin{widetext}
\ba
\nonumber
&&\left(\int dt\,\epsilon_{A}(t) (e^{i \Omega \tau_{A}(t)}\sigma_A^{+}+e^{-i \Omega \tau_{A}(t) }\sigma_A^{-}) \otimes {\phi}_{A}(t)\right)\rho_{AB}(0)\otimes\dyad{0}_\phi\left(\int dt'\,\epsilon_{A}(t') (e^{i \Omega \tau_{A}(t')}\sigma_A^{+}+e^{-i \Omega \tau_{A}(t') }\sigma_A^{-}) \otimes {\phi}_{A}(t')\right)\\
\nonumber
&&=\left(\sigma_A^{+}\phi^{+}(t)+\sigma_A^{-}\phi^{-}(t)\right)\rho_{AB}(0)\otimes\dyad{0}_\phi\left(\sigma_A^{+}\phi^{+}(t')+\sigma_A^{-}\phi^{-}(t')\right)\\
\nonumber
&&=\sigma_A^{+}\rho_{AB}(0)\sigma_A^{+}\otimes\dyad{E_A^+}+\sigma_A^{+}\rho_{AB}(0)\sigma_A^{-}\otimes\dyad{E_A^+}{E_A^{-}}+\sigma_A^{-}\rho_{AB}(0)\sigma_A^{+}\otimes\dyad{E_A^{-}}{E_A^+}+\sigma_A^{-}\rho_{AB}(0)\sigma_A^{-}\otimes\dyad{E_A^-}\\
&&\nonumber
\ea
\end{widetext}

Where we have defined $\epsilon_{i}(t)=\epsilon(\tau_i(t))\frac{d\tau_i}{dt}$ and introduced the notation:
\ba
\nonumber
&&\\ &&\phi_i(t)=\phi(\bm{x}_i(t))\\
&&{\phi_i}^{\pm}=\int dt \epsilon_i(t) e^{\pm i \Omega \tau_i(t)} {\phi}_i\\
\nonumber
&&\ket{E_i^{\pm}}={\phi}_i^{\pm}\ket{0}_{\phi} 
\ea

At this point one can trace away the field degrees of freedom obtaining:
\begin{widetext}
\ba
\nonumber
&&\Tr_\phi[\left(\int dt \frac{d\tau_A}{dt} {H}_{\rm int}^{(A)} (\tau_A(t))\right)\rho(0)\left(\int dt' \frac{d\tau_A}{dt'} {H}_{\rm int}^{(A)} (\tau_A(t'))\right)]\\
\nonumber
&&=\braket{E_A^+}\sigma_A^{+}\rho_{AB}(0)\sigma_A^{+}+\braket{E_A^+}{E_A^{-}}\sigma_A^{+}\rho_{AB}(0)\sigma_A^{-}+\braket{E_A^{-}}{E_A^+}\sigma_A^{-}\rho_{AB}(0)\sigma_A^{+}+\braket{E_A^-}\sigma_A^{-}\rho_{AB}(0)\sigma_A^{-}
\ea
\end{widetext}
where the inner products between states of the field are worth:
\ba
\nonumber
\braket{E_i^{\pm}}{E_j^{\pm}}=\int dt\,dt'\,\epsilon_i(t)\epsilon_j(t')e^{i\Omega(\pm t\pm t')}W(t,t')
\ea
and we have defined the Wightman functions:
\ba
\nonumber
W(t,t')=\bra{0}\phi(t)\phi(t')\ket{0}.
\ea
At this point one has to realize that all other possible combinations of $H_{\rm int}^{(A)}$ and $H_{\rm int}^{(B)}$ will give rise to similar terms, and that the same holds also for terms stemming from the trace of the second order operator $U_2$. Once this is realized, it is easy to see, after some standard algebra, that the initial state of the detector is mapped into the final one via:
\ba
\nonumber
&&\mathcal{E}(\rho_{AB}(0))=\rho_{AB}(0)+\sum_{i,j=A,B}\sum_{\alpha,\beta=\pm}\\
\nonumber
&&\braket{E_i^\alpha}{E_j^\beta}\left(\sigma_i^\alpha\rho_{AB}(0)\sigma_j^\beta-\frac{1}{2}\acomm{\sigma_j^\beta\sigma_i^\alpha}{\rho_{AB}(0)}\right)
\ea

For instance, when the detectors are initialized in the state $\ket{00}$, the final detectors density matrix $\rho_{AB}$ is given by:
\ba
\label{eq:final_state_00}
\begin{pmatrix}
1-\lvert E_A \rvert^2-\lvert E_B \rvert^2  & 0 & 0& M \\
0& \lvert E_B\rvert^2  & \bra{E_A^-}\ket{E_B^+} &0 \\
0&\bra{E_B^-}\ket{E_A^+} &\lvert E_A \rvert^2 &  0 \\
M^*&0&0&0 
\end{pmatrix}\;
\ea 
where $\absval{E_i}^2=\braket{E_i^-}{E_i^+}$ is the probability of having the $i$-th detector excited after the interaction, $\bra{E_B^+}\ket{E_A^-}$ is the overlap between the excited states of the two detectors and $M=-\left(\braket{E_B^-}{E_A^-}+\braket{E_A^-}{E_B^-}\right)/2$ is the probability that the two detectors exchange a virtual particle. Once the parameters of the interaction are fixed, all terms only depend on the trajectories of the detectors, whose information is contained in the Wightman functions.  The reader can look also at Appendix~\ref{sec:app_den_mat} for a detailed derivation of the final state of the detectors  in Eq.~\eqref{eq:final_state_00} and at Appendix~\ref{sec:other_states} for the expression of other final states when the detectors are initialized in different states used in Sec.~\ref{sec:accel}.

\subsection{Perturbative calculation of the Concurrence}
We now want to show the perturbative  computation of the concurrence for the final state of the detectors when they are initialized in the state $\ket{00}_{AB}$. 
Let us first report also here the formula of the concurrence $\mathcal{C}$:
\ba
\nonumber
\mathcal{C}(\rho)=\max\{0,\lambda_1-\lambda_2-\lambda_3-\lambda_4\}
\ea
with the $\lambda_i$ being the square roots of the eigenvalues of the matrix $R=\rho\tilde{\rho}$.
In order to compute the concurrence, we first need to compute the matrix $\tilde{\rho}=(Y\otimes Y)\rho^*(Y\otimes Y)$.
For the final state of the detectors in Eq.~\eqref{eq:final_state_00}, this matrix can be written formally as:
\ba
\nonumber
\tilde{\rho}=\begin{pmatrix}
\rho_{44}&0&0&\rho_{14}\\
0&\rho_{33}&\rho_{23}&0\\
0&\rho_{23}^*&\rho_{22}&0\\
\rho_{14}^*&0&0&\rho_{11}
\end{pmatrix}
\ea
Notice that we have reintroduced the element $\rho_{44}$: while this is of order $\lambda^4$, it will enter the expression for the eigenvalues of $\rho\tilde{\rho}$ under a square rot, leading to a contribution of order $\lambda^2$ that shall not be neglected. Indeed, this matrix element can be evaluated using the same techniques of App.~\ref{sec:app_den_mat}, obtaining
\ba
\nonumber
\rho_{44}=|M|^2+|\langle E_A^-|E_B^+\rangle|^2+|E_A|^2|E_B|^2+\mathcal{O}(\lambda^6)
\ea
see~\cite{detectors} for further details.

At this point we can compute the square root of the eigenvalues of the matrix $R$, obtaining:
\ba
\nonumber
&&\lambda_1=\sqrt{\rho_{11}\rho_{44}}+|\rho_{14}|=\\
\nonumber
&&\sqrt{|M|^2+|\langle E_A^-|E_B^+\rangle|^2+|E_A|^2|E_B|^2}+|M|+\mathcal{O}(\lambda^4)\\
\nonumber
&&\lambda_2=\sqrt{\rho_{22}\rho_{33}}+|\rho_{23}|=\\
\nonumber
&&\sqrt{|E_A|^2|E_B|^2}+|\langle E_A^-|E_B^+\rangle|+\mathcal{O}(\lambda^4)\\
\nonumber
&&\lambda_3=\sqrt{\rho_{11}\rho_{44}}-|\rho_{14}|=\\
\nonumber
&&\sqrt{|M|^2+|\langle E_A^-|E_B^+\rangle|^2+|E_A|^2|E_B|^2}-|M|+\mathcal{O}(\lambda^4)\\
\nonumber
&&\lambda_4=\sqrt{\rho_{22}\rho_{33}}-|\rho_{23}|=\\
\nonumber
&&\sqrt{|E_A|^2|E_B|^2}-|\langle E_A^-|E_B^+\rangle|+\mathcal{O}(\lambda^4)
\ea
First, let us notice that only terms up to the order $\mathcal{O}(\lambda^2)$ appear in the final expression. Notice also the matrix element $\rho_{44}$ under the square root sign in the expressions of $\lambda_{1,3}$: when multiplied by the zeroth order term in $\rho_{11}$, the contributions from $\rho_{44}$ are of second order after one takes the square root.
Finally, following~\cite{detectors}, we notice that the greatest of these eigenvalues, when the two detectors are entangled, is just $\lambda_1$, so that the  concurrence can be finally written as:
\ba
\nonumber
\mathcal{C}=\max\{0,2(|M|-\sqrt{|E_A|^2|E_B|^2})\}+\mathcal{O}(\lambda^4)
\ea
Naturally, the same procedure can be repeated also for other final states of the detectors following the same procedure.

\subsection{Perturbative calculation of the SRE\label{sec:perturbative_SRE}}

At this point, we are left with the calculation of the SRE. We will work out the calculation for the state in Eq.~\eqref{eq:final_state_00}, but the procedure can be easily generalized to any other state.
Starting from the definition in Eq.~\eqref{eq:sre_mixed}, we need to compute the term $M_2(\rho)=\Tr[Q\rho^{\ot4}]$. For a 2-qubit system one has  $Q:=d^{-2}\sum_{P\in\pauli{2}}P^{\otimes4}$ and $d=4$.
The set of 2-qubit Pauli strings includes the sixteen $4\times 4$ matrices obtained by all possible combinations of the tensor product of two Pauli operators, namely:
\ba
\nonumber
\mathbb{P}_2=\{&&\Id\ot\Id , \Id\ot X, \Id \ot Y, \Id \ot Z,\\
\nonumber
&&X \ot \Id, X \ot X, X \ot Y, X \ot Z,\\
\nonumber
&& Y \ot \Id, Y \ot X, Y\ot Y, Y\ot Z,\\
&&Z \ot \Id, Z \ot X, Z \ot Y, Z \ot Z\}.
\ea
Using the properties of the trace, one can rewrite Eq.~\eqref{eq:sre_mixed} as:
\ba 
\nonumber
M_2(\rho)=-\log \sum_{P\in\mathbb{P}_2}\frac{\Tr\{P\rho\}^4}{d} + \log \Tr{\rho^2}
\ea
To calculate the non-stabilizerness of the detectors state $\rho_{AB}$ in~\eqref{eq:final_state_00}, one needs to sum the fourth powers of the traces of the product between the density matrix and the $16$ elements of the Pauli group.
To do this we follow the same approach used for the concurrence, namely we first compute the traces using a formal expression of the density matrix, and then we compute the fourth power of these expressions and cut all terms beyond the order $\lambda^2$.
\begin{widetext}
Assuming a matrix of the form shown in~\eqref{eq:final_state_00} one obtains the following non zero contributions:
\ba
\nonumber
\Tr{(\Id\ot\Id) \rho_{AB}} &=&\rho_{11}+\rho_{22}+\rho_{33}+\rho_{44}= 1+\mathcal{O}(\lambda^4) \\
\nonumber
\Tr{(\Id\ot Z) \rho_{AB}} &=&\rho_{11}-\rho_{22}+\rho_{33}-\rho_{44} =1-2\absval{E_B}^2 +\mathcal{O}(\lambda^4)\\
\nonumber
\Tr{(X \ot X) \rho_{AB}} &=&\rho_{14}+\rho_{14}^*+\rho_{23}+\rho_{23}^*= \bra{E_A}\ket{E_B} + \bra{E_A}\ket{E_B}^* + M + M^* +\mathcal{O}(\lambda^4)\\
\nonumber
\Tr{(X\ot Y) \rho_{AB}} &=&i(\rho_{14}-\rho_{14}^*+\rho_{23}-\rho_{23}^*) = i(M-M^* - \bra{E_A}\ket{E_B} + \bra{E_A}\ket{E_B}^*) +\mathcal{O}(\lambda^4)\\
\nonumber
\Tr{(Y\ot X) \rho_{AB}} &=&(\rho_{14}-\rho_{14}^*-\rho_{23}+\rho_{23}^*) = i(M-M^* + \bra{E_A}\ket{E_B} - \bra{E_A}\ket{E_B}^*) +\mathcal{O}(\lambda^4)\\
\nonumber
\Tr{(Y\ot Y) \rho_{AB}} &=&-\rho_{14}-\rho_{14}^*+\rho_{23}+\rho_{23}^* = -M-M^* + \bra{E_A}\ket{E_B} + \bra{E_A}\ket{E_B}^*) +\mathcal{O}(\lambda^4)\\
\nonumber
\Tr{(Z\ot \Id) \rho_{AB}} &=&\rho_{11}+\rho_{22}-\rho_{33}-\rho_{44}  = 1-2\absval{E_A}^2+\mathcal{O}(\lambda^4) \\
\nonumber
\Tr{(Z\ot Z) \rho_{AB}} &=&\rho_{11}-\rho_{22}-\rho_{33}+\rho_{44}  = 1 - 2 \absval{E_A}^2 - 2\absval{E_B}^2 +\mathcal{O}(\lambda^4).
\ea 
\end{widetext}
Similarly to the case of the concurrence, one has then to take the fourth power of all these contributions, and then cut all terms beyond $\mathcal{O}(\lambda^2)$,obtaining:
\ba
\nonumber
\Tr{(\Id\ot\Id) \rho_{AB}}^4 &=& 1 \\
\nonumber
\Tr{(\Id\ot Z) \rho_{AB}}^4 &=& 1-8\absval{E_B}^2 \\
\nonumber
\Tr{(X \ot X) \rho_{AB}}^4 &=& 0\\
\nonumber
\Tr{(X\ot Y) \rho_{AB}}^4 &=& 0\\
\nonumber
\Tr{(Y\ot X) \rho_{AB}}^4 &=& 0\\
\nonumber
\Tr{(Y\ot Y) \rho_{AB}}^4 &=& 0\\
\nonumber
\Tr{(Z\ot \Id) \rho_{AB}}^4 &=& 1-8\absval{E_A}^2 \\
\nonumber
\Tr{(Z\ot \Id) \rho_{AB}}^4 &=& 0 \\
\nonumber
\Tr{(Z\ot Z) \rho_{AB}}^4 &=& 1 - 8 \absval{E_A}^2 - 8\absval{E_B}^2
\ea
where all the above quantities are meant up to $\mathcal{O}(\lambda^2)$.
One can then sum up all the contributions above to get the expression for the 2-SRE of the $\rho_{AB}$ in Eq.~\eqref{eq:final_state_00}:
\ba
\nonumber
 M_2(\rho_{AB})&=& -\log [1-4(|E_A|^2+|E_B|^2)]\\
 \nonumber
 &+&\log [1-2(|E_A|^2+|E_B|^2)]
 \ea 
where the first logarithm stems from the term proportional to the fourth power of the expectations value over the Pauli operators, while the second logarithm is just (minus) the 2-Rényi entropy of the detectors state.

\subsection{Perturbative CHSH inequality}
Let us finally show the perturbative expression of the expectation value of the Bell operator in Eq.~\eqref{eq:bell_operator}, for the final state of the detector in Eq.~\eqref{eq:final_state_00}. As the expectation value of the Bell operator is linear in the density matrix elements, one can simply shove the perturbative final state into the trace. Taking the trace of $\rho_{AB}$ and the various Pauli strings in the Bell operator $\mathcal{B}_0$ we obtain:
\ba
\nonumber
&&\Tr\left[\rho_{AB}(X_A\otimes X_B)\right]=\rho_{14}+\rho_{41}+\rho_{23}+\rho_{32}\\
\nonumber
&&=\bra{E_A}\ket{E_B} + \bra{E_A}\ket{E_B}^* + M + M^* +\mathcal{O}(\lambda^4)\\
\nonumber
&&\Tr\left[\rho_{AB}(X_A\otimes Z_B)\right]=0\\
\nonumber
&&\Tr\left[\rho_{AB}(Z_A\otimes X_B)\right]=0\\
\nonumber
&&\Tr\left[\rho_{AB}(Z_A\otimes Z_B)\right]=\rho_{11}-\rho_{22}-\rho_{33}+\rho_{44}\\
\nonumber
&&= 1 - 2 \absval{E_A}^2 - 2\absval{E_B}^2 +\mathcal{O}(\lambda^4)
\ea
Putting everything together we obtain:
\ba
\nonumber
\Tr\left[\mathcal{B}_0\rho_{AB}\right]&=&1 - 2 \absval{E_A}^2 - 2\absval{E_B}^2+\bra{E_A}\ket{E_B}\\
\nonumber
&&+ \bra{E_A}\ket{E_B}^* + M + M^* +\mathcal{O}(\lambda^4)
\ea

A similar expression can also be found for the case of initially entangled detectors, see Appendix~\ref{sec:other_states} for further details. 

\section{Results\label{sec:results}}

\subsection{The inertial scenario}
\begin{figure}[!t]
\centering
\includegraphics[width=\linewidth]{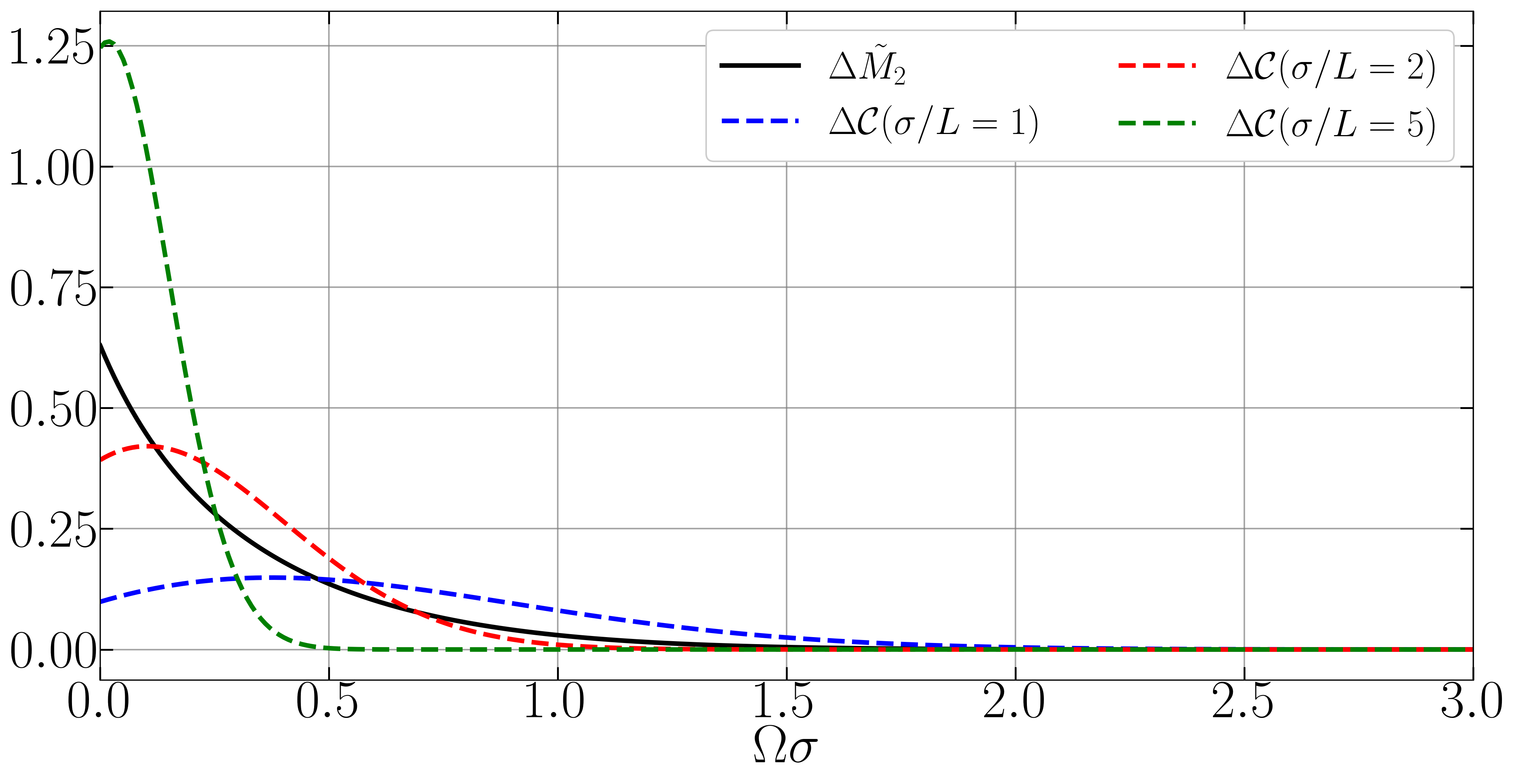}
\caption{Plot of the SRE as a function of $\Omega\sigma$ and of the concurrence for different values of $\sigma/L$. The initial state of the detectors is $\rho_{AB}=\dyad{00}$. For small values of the product $\Omega\sigma$ one can see that the final state of the detectors has significant amount of non-stabilizerness. While this is in contrast with the vacuum state of the field being a stabilizer state (i.e. SRE-free state), the presence of SRE is explained by noticing that the interaction between the detectors and the field is non-Clifford, such that some non-stabilizerness is introduced in the system for small values of $\sigma\Omega$. On the other hand, for $\Omega\sigma\gtrsim1$, one can get rid of this spurious contribution and genuinely harvest SRE from the field.}
\label{fig:inertial}
\end{figure}

Let us start our analysis from the simpler case of an inertial setting, with the detectors initialized in the state $\rho_{AB}(0)=\dyad{00}$. In this case $t=\tau_i$, and the Wightman function of the field can be computed analytically~\cite{detectors} as:
\ba
\nonumber
W(t,t')=\frac{1}{4\pi i}\,\rm{sgn}(t-t')\delta(f(t,t'))-\frac{1}{4\pi^2f(t,t')}\\
\ea
where
\ba
\nonumber
f(x,x')=(t-t')^2-L^2.
\ea
$L$ being the initial separation between the detectors.
We can exploit this result to obtain an analytical expression for the transition probabilities $|E_A|^2,|E_B|^2$ on which both the concurrence and the SRE depend when the detectors are initialized in the state $\ket{00}_{AB}$. We choose a switching function of the Gaussian form $\epsilon_i(t)=e^{-\frac{\tau_i^2}{2\sigma^2}}$, obtaining for the transition probabilities:
\ba
\label{eq:inertial_transition}
|E_i|^2=\frac{\lambda^2}{4\pi}\left[e^{-\sigma^2\Omega^2}-\sqrt{\pi}\sigma\Omega\rm{erfc}(\sigma\Omega)\right].
\ea
One can immediately notice that in this expression only the frequency of the detectors $\Omega$ and the width of the switching function $\sigma$ play a role, so that the SRE will only depends upon these parameters. We plot in Fig.~\ref{fig:inertial} the SRE as a function of the frequency $\Omega$ for various values of $\sigma$. One can observe that the SRE decreases quickly when $\Omega$ becomes larger than $\sigma$, becoming zero already for $\Omega\simeq2\sigma$. Also, noticeably the SRE does not depend on the initial separation $L$ between the detectors.

As for the concurrence, one has to compute also the matrix element $M$, which in the inertial setting can be also analytically computed as:
\ba
\nonumber
M=i\frac{\lambda^2\sigma}{4\sqrt{\pi}L}e^{-\sigma^2\Omega^2-\frac{L^2}{4\sigma^2}}\left[\rm{erf}\left(i\frac{L}{2\sigma}\right)-1\right]
\ea
One can immediately observe that in contrast with SRE, the concurrence does depend on the initial separation $L$ between the two detectors.

To analyze the quantum resources of the evolved state, we normalize these quantities with respect to $\lambda^2$. This choice ensures a well-defined comparison across different parameters regimes and captures the leading-order behavior in the perturbative expansion. Moreover, we study the variation of both SRE and concurrence:
\ba
\nonumber
\Delta\tilde{M}_2(\rho_{AB})=\tilde{M}_2(\mathcal{E}(\rho_{AB}))-M_2(\rho_{AB}(0))
\ea
and similarly for the concurrence. Notice that in the inertial case under examination this is just equivalent to computing the concurrence and SRE of the final state, since initially both resources are zero in the detectors state. This will not be the case when studying the accelerated scenario with different initial states, see Sec.~\ref{sec:accel}.

Let us now comment on the results in the plots in Fig.~\ref{fig:inertial}. One can observe that for small values of $\sigma\Omega$, the final state of the detectors has a significant amount of SRE, which however goes quickly to zero for larger values of $\sigma\Omega$. A similar behavior is observed for the concurrence. While the behavior of the concurrence has been largely analyzed in the literature on entanglement harvesting, let us focus on the behavior of the SRE. 
The interpretation of this results is easily given: while in an inertial reference frame the field is in a stabilizer state, so that no resource is expected to be harvested, the interaction between the detectors and the field is actually non-Clifford, so that some non-stabilizerness is introduced in the system. Thus, in an inertial setting, all non-stabilizerness is spurious, rather than harvested from the field. However, this effect becomes negligible for large enough values of $\sigma\Omega$. This will justify our choice of parameters when studying the accelerated scenario, where no analytical expressions can be obtained: by choosing $\sigma\Omega=2$ we will ensure that all the harvested SRE will stem from the acceleration, without contributions due to the non-Clifford interaction between the detectors and the field. 

\subsection{Accelerated detectors: the parallel scenario\label{sec:accel}}

\begin{figure*}[!t]
\centering
\begin{subfigure}{0.49\textwidth}
\centering
\includegraphics[width=\linewidth]{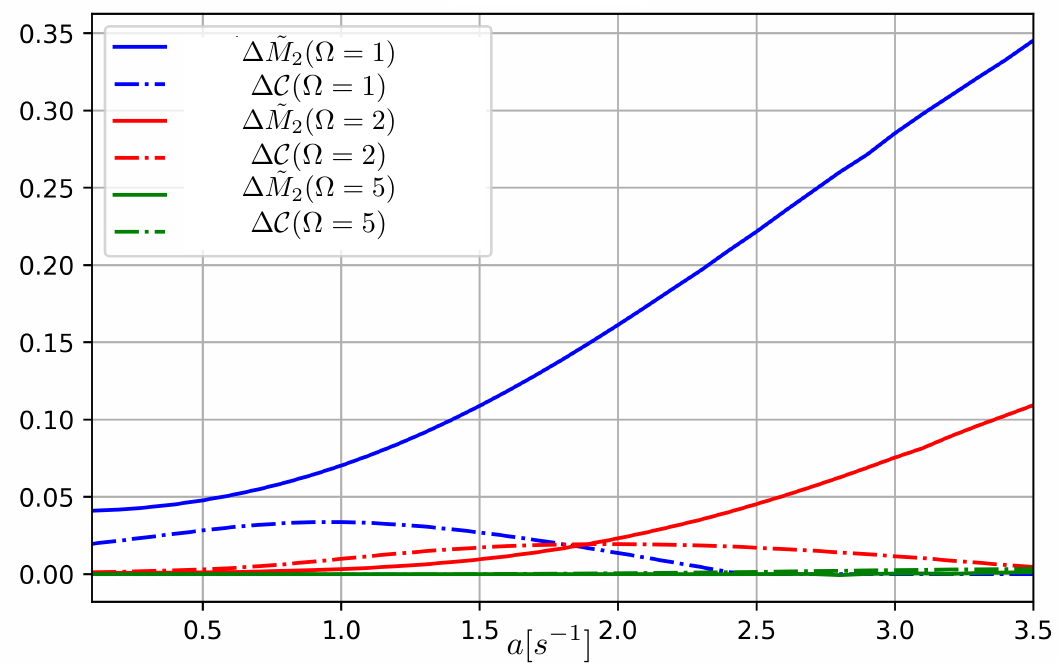}
\caption{}
\label{fig:parameters_omega}
\end{subfigure}
\begin{subfigure}{0.49\textwidth}
\centering
\includegraphics[width=\linewidth]{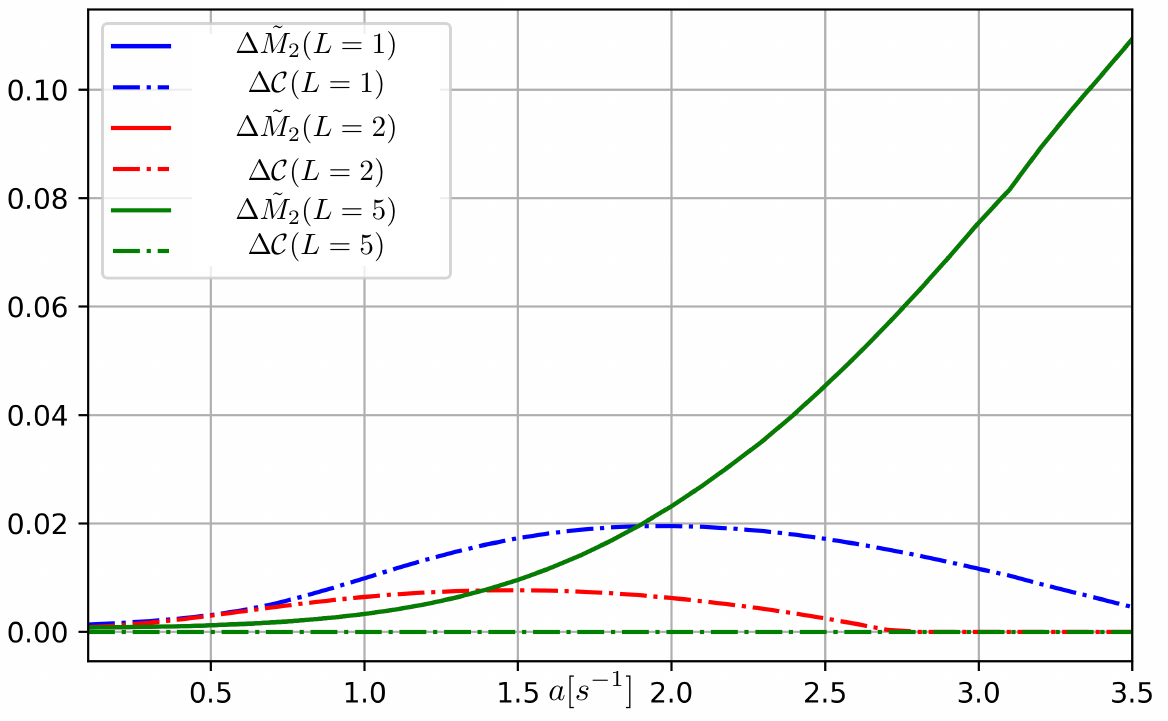}
\caption{}
\label{fig:parameters_L}
\end{subfigure}
\caption{Variations of SRE $\Delta\tilde{M}_2(\rho_{AB})$ (solid line) and concurrence  $\Delta\mathcal{C}(\rho_{AB})$ (dashed line), as a function of the (parallel) acceleration $a_\parallel$, for various choices of parameters, with the detectors initialized in the state $\rho_{AB}=\dyad{00}$. In Fig.~\ref{fig:parameters_omega} we study the two resources for different values of $\Omega$, setting $\sigma=1$ and $L=1$. Notice that for $\Omega=1$, for $a=0$ one has a non zero SRE, in agreement with the observation made in the inertial setting. One can also notice that  the harvested SRE becomes larger for lower values of $\Omega$. In Fig.~\ref{fig:parameters_L} we plot the resources for different values of the initial separation $L$ between the detectors, setting $\Omega=2$ and $\sigma=1$. Also in this case, one can observe the same behavior obtained in  the inertial setting, as the SRE does not depend on $L$, while the concurrence does.}
\label{fig:parameters}
\end{figure*}

\begin{figure*}[!t]
\centering
\begin{subfigure}{0.33\textwidth}
\centering
\includegraphics[width=\linewidth]{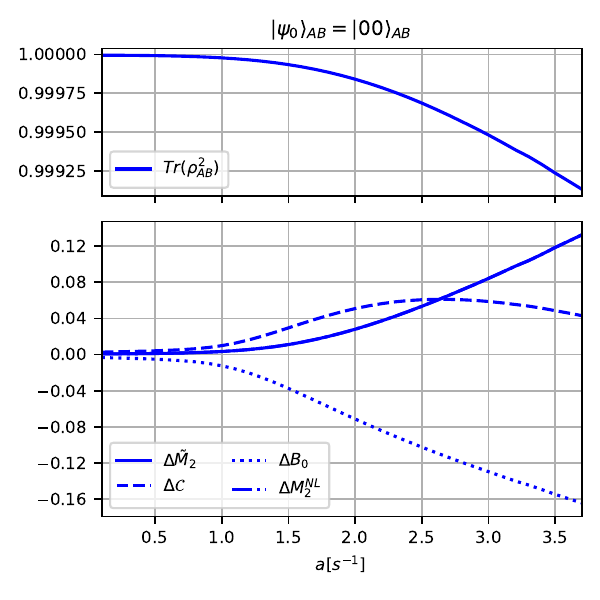}
\caption{}
\label{fig:00_par}
\end{subfigure}
\begin{subfigure}{0.33\textwidth}
\centering
\includegraphics[width=\linewidth]{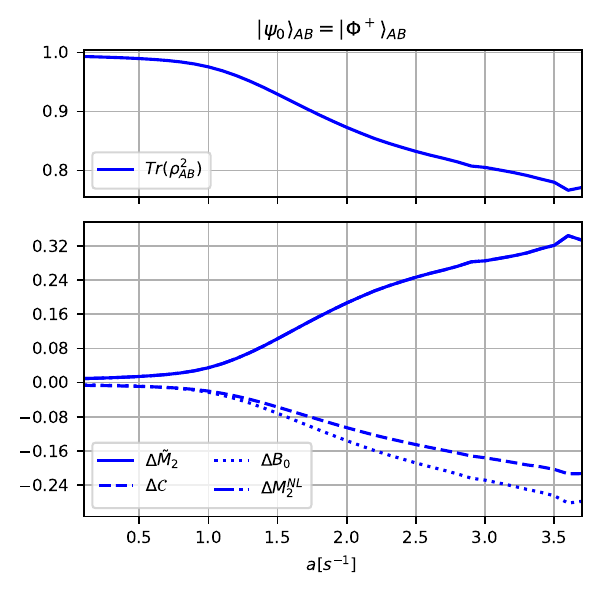}
\caption{}
\label{fig:bell_par}
\end{subfigure}
\begin{subfigure}{0.33\textwidth}
\centering
\includegraphics[width=\linewidth]{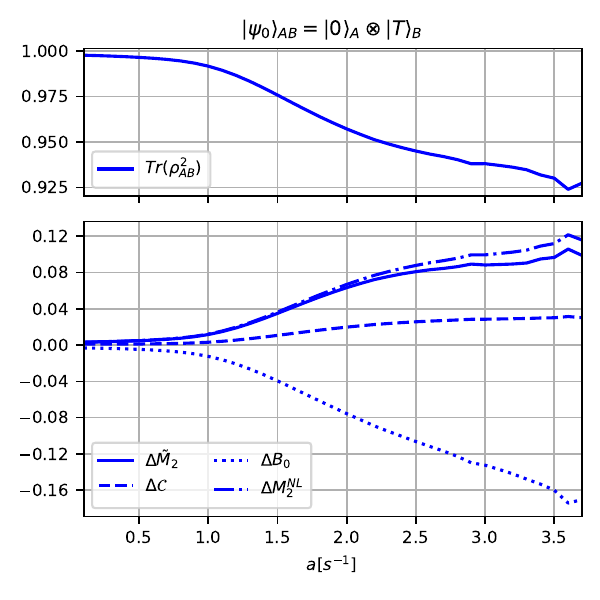}
\caption{}
\label{fig:0T_par}
\end{subfigure}
\caption{ Variations of SRE $\Delta\tilde{M}_2(\rho_{AB})$ (solid line), concurrence  $\Delta\mathcal{C}(\rho_{AB})$ (dashed line), non-local SRE $\Delta M^{NL}_2 (\rho_{AB})$ (dashed-dotted line) and value of Bell operator $\Delta B_0$ (dotted line) as a function of the (parallel) acceleration $a_\parallel$, together with the corresponding purity of the state (see upper plots). Panels (a) Initial state $\ket{00}$, (b) initial state $\ket{\Phi^+}$, (c) initial state $\ket{0}_A\otimes\ket{T}_B$. In panel (a) and (b) the curves for $\Delta\tilde{M}_2(\rho_{AB})$  and $\Delta M^{NL}_2 (\rho_{AB})$ overlap perfectly as all the harvested SRE is non-local. Notice that the state of the detectors, for the initial state $\ket{\psi_0}_{AB}=\ket{00}$, is always pure with very good approximation.
}
\label{fig:plots}
\end{figure*}

Let us now move to the accelerated setting. We start our analysis by studying how the harvested SRE varies with the physical parameters describing the system, namely the frequency $\Omega$ of the two detectors, the interaction time $\sigma$ and the initial separation $L$ between the two probes. We will focus our analysis on the case of parallel acceleration.
First of all, in the case of accelerated detectors, the Wightman functions are not analytically solvable, because of the hyperbolic functions in the worldlines of the detectors (see App.~\ref{sec:app_accel}). We thus rely on a numerical analysis, as shown in Figs.~\ref{fig:parameters},\ref{fig:plots}.

From the plots in Fig.~\ref{fig:parameters}, for the detectors initialized once again in the state $\rho_{AB}(0)=\dyad{00}$,  we can observe some common features with the case of inertial reference frames. In Fig.~\ref{fig:parameters_omega} we can observe that for small values of $\sigma\Omega$ there is a contribution to the SRE due to the interaction being non Clifford. This effect is once again negligible for large enough values of $\sigma\Omega$. In Fig.~\ref{fig:parameters_L} we once again observe the independence of the SRE from the initial distance between the detectors.

Let us now turn our attention on the nature of the harvested SRE. Specifically, we now want to focus on local and non-local SRE, their interplay with the concurrence when varying the intial state of the detectors.
As anticipated when discussing the inertial setting, we set the interaction parameters to $\Omega=2$, $\sigma=1$ and $L=0.5$, in order to avoid spurious contributions given by the non-Clifford interaction between the detectors and the field, and be sure  that all the SRE is effectively harvested from the field. Also in this case we focus here on the parallel scenario, leaving the antiparallel and perpendicular cases in Appendix~\ref{sec:app_accel}.

Starting from  different initial states of the detectors, we compute the
variation of SRE $\Delta\tilde{M}_2(\rho_{AB})$, of the concurrence $\Delta\mathcal{C}(\rho_{AB})$, of the expectation value of the Bell operator in Eq.~\eqref{eq:bell_operator} $\Delta B_0$, and of the non-local SRE $\Delta M^{NL}_2 (\ket{\psi})$ with respect to their initial state values.
Specifically, we consider three different initial states $\ket{\psi_0}_{AB}$ of the detectors: the resource-free state $\ket{00}_{AB}$,  the maximally entangled, SRE-free state $\ket{\Phi^+}_{AB}$ and the separable, SRE-full state $\ket{0}_A\otimes\ket{T}_B$ with $\ket{T}=(\ket{0}+e^{i\pi/4}\ket{1})/\sqrt{2} $.

\begin{table}[!t]
\centering
\begin{tabular}{|c|c|c|c|c|}
\toprule
$\rho_{AB}(0)$&$\tilde{M}_2$&$\mathcal{C}$&$b_0$ & $M^{NL}_2$\\
\midrule
$\ket{00}$&$0$&$0$&$1$ &$0$\\
$\ket{\Phi^+}$&$0$&$1$&$2$ &$0$\\
$\ket{0}\otimes\ket{T}$&$\simeq 0.415$&$0$&$-\frac{\sqrt{2}}{2}$ &$0$\\
\bottomrule
\end{tabular}
\caption{Initial values of SRE $\tilde{M}_2$, concurrence $\mathcal{C}$, Bell operator $b_0$ and non-local SRE $M^{NL}_2$ for different choices of the initial state of the detectors $\rho_{AB}(0)$.}
\label{tab:initial_condition}
\end{table}

We summarize the initial values of the quantities of interest for the different initial states in Table~\ref{tab:initial_condition}. We compute the variations $\Delta\tilde{M}_2(\rho_{AB})$, $\Delta\mathcal{C}(\rho_{AB})$, $\Delta B_0$, $\Delta M^{NL}_2 (\ket{\psi})$ of our quantities
as a function of the acceleration $a$ for the three choices of initial states, see Fig.~\ref{fig:plots}.

In all cases, one harvests both entanglement $\mathcal{C}$ and SRE $\tilde{M}_2$. 
We see that the harvesting of resources is non-monotone with the acceleration $a_\parallel$. Remarkably, if there is no SRE in the initial state, {\em all} the harvested SRE is non-local, so this is genuine harvesting in the strong sense. Even if there is some local SRE in the initial state, most of the harvested SRE is still non-local, see fig.\ref{fig:0T_par}. In this latter case, even the local part of the
harvested SRE would not be produced if the acceleration were zero, and the field would be in a stabilizer state. At finite accelerations, the SRE in the field enables the production of SRE in the detectors: this is harvesting in the weak sense\cite{footnote}.

\subsection{Non-local violations}

\begin{figure*}[!th]
\centering
\begin{subfigure}{0.49\textwidth}
\centering
\includegraphics[width=\linewidth]{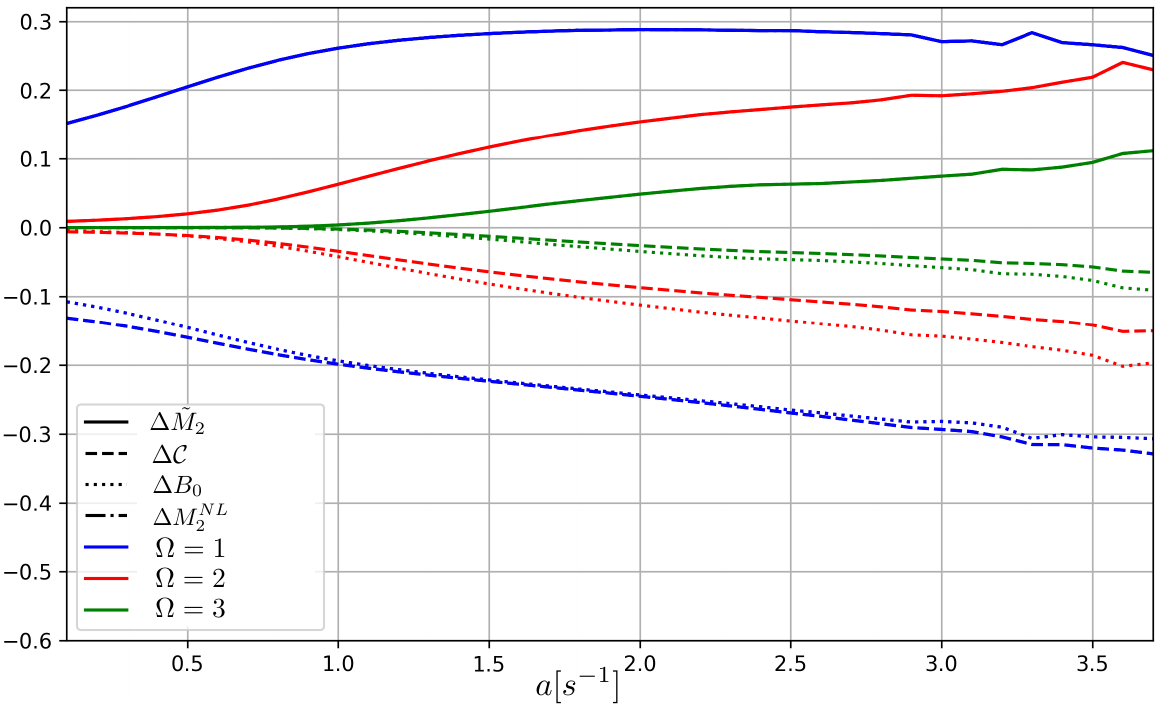}
\caption{}
\label{fig:bell_acc_omega}
\end{subfigure}
\begin{subfigure}{0.49\textwidth}
\centering
\includegraphics[width=\linewidth]{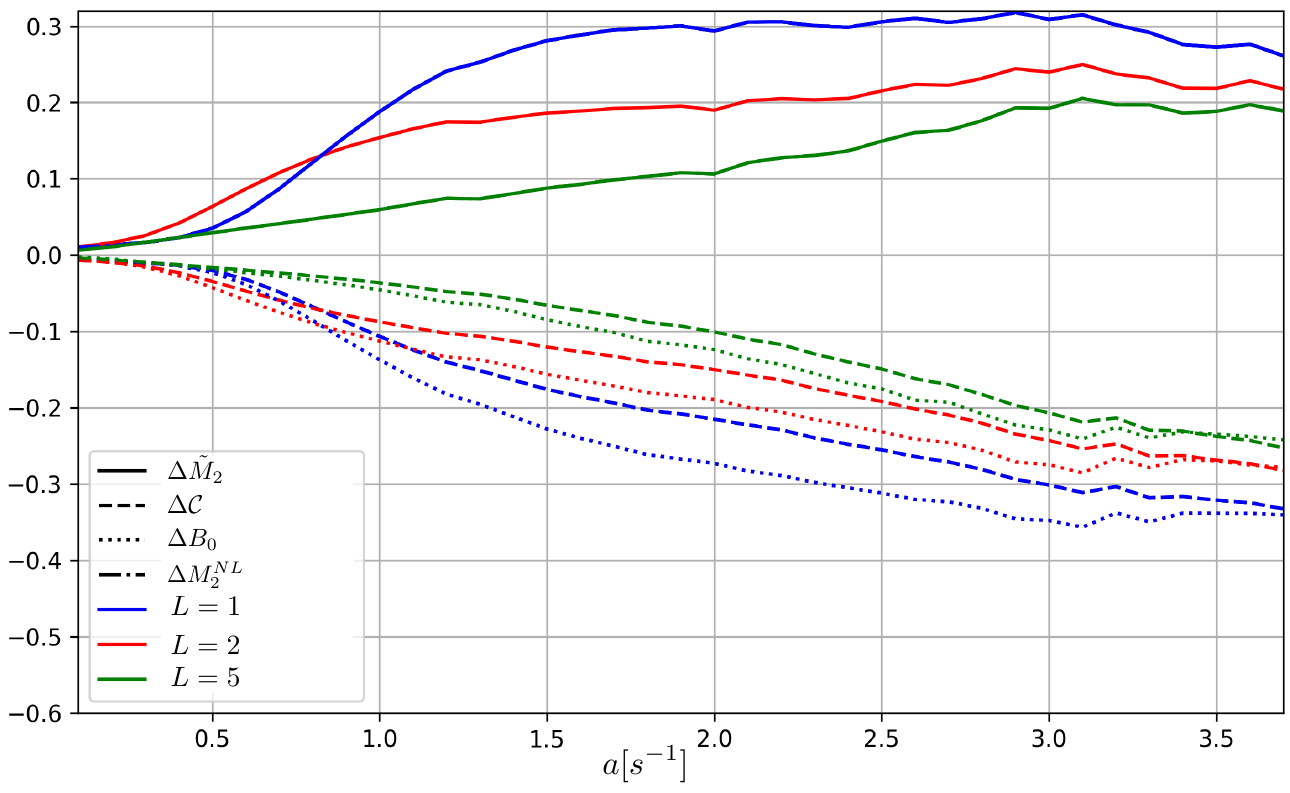}
\caption{}
\label{fig:bell_acc_L}
\end{subfigure}
\caption{Plot of the variation of the non-stabilizerness, concurrence, expectation value of the Bell operator and non-local non-stabilizerness as a function of the acceleration when the detectors are initialized in the maximally entangled state $\rho_{AB}(0)=\dyad{\Phi^+}$ for different values of the detectors frequency. In the plot we have set $\sigma=1$ and $L=1$. Different line shapes represent the various quantities, while different colours represent different values of $\Omega$, as in the legend. From a qualitative point of view, one can once again observe that all the harvested non-stabilizerness is non-local. This feature, as explained in the main text, is the one responsible for the non violation of the CHSH inequality. The only difference with respect to the scenario where the detectors are initialized in the state $\dyad{00}$ is that the SRE also depends on the initial separation $L$. However, there are no choice of parameters for which the CHSH inequality is violated.}
\label{fig:bell_acc_parameters}
\end{figure*}

Importantly, in all cases the CHSH is not violated.
We remark here that similar results are obtained, especially regarding CHSH, for any choice of Pauli measurement in the definition of the Bell operator $\mathcal{B}_0$, Eq.~\eqref{eq:bell_operator}. If one starts with resource-free states, there is not enough harvesting of the resources to guarantee a CHSH violation. Furthermore, if one saturates one of the resources, e.g., entanglement or SRE (on one of the detectors), the interaction with the field causes a loss of the resource as the other one is harvested. This behavior is confirmed for different choices of parameters, as shown Fig.~\ref{fig:bell_acc_parameters}, where for different values of $\Omega$ one can observe that the CHSH inequality is never violated. In a way or another, the form of non-locality displayed by CHSH violations cannot be induced by the interaction with the field. 

Some explanations are in order: one can harvest substantial amounts of SRE and entanglement, and even non-local SRE. However,  it is not possible to harvest non-locality in the form of CHSH violation from the quantum field. What is at play here? There are two main elements to this result. First, if all the SRE is non-local, there cannot be any violation of the CHSH inequalities, as shown in~\cite{cusumano2025nonstabilizernessviolationschshinequalities}. Moreover, the nature of the harvested SRE does not depend on the physical parameters of the system, such as $\Omega, L, \sigma$, but rather on the form of the interaction between the detectors and the field, so that no violation of the CHSH inequality would be observed for any choice of these parameters. For a two-qubit system, (pure) states possessing only non-local non-stabilizerness are completely determined by their Schmidt coefficients as $\sqrt{x} \ket{00} +\sqrt{1-x}\ket{11}$, see \cite{cao2024gravitationalbackreactionmagical, qian2025quantumnonlocalmagic} and it is straightforward to check that  such states cannot violate the CHSH inequalities for $B_0$. So, as one only harvests non-local SRE, non-locality effects lack the necessary resources. In our setting, at least for the initial state $\ket{00}$ the time evolution, in spite of the system being open, keeps the state approximately pure, see the upper plot in Fig.\ref{fig:00_par} and these conclusions hold. But there is more: one could in fact engineer detectors that couple with a field in a different way, so that also local non-stabilizer resources are injected. However, homogeneity of space imposes that such process should be symmetric for the exchange of the two detectors. One can prove that \cite{cusumano2025nonstabilizernessviolationschshinequalities}
the only way of violating CHSH inequalities is to prepare the state in an asymmetric way, so such violations would be impossible in principle for identical detectors.

\section{Conclusions and outlook}
In this paper, we have shown a protocol for SRE harvesting in both inertial and accelerated reference frames, highlighting its main features and dependencies of the physical parameters. The main result of the paper is that SRE can be harvested from the vacuum state of a relativistic quantum field and stored into the state of a detector interacting with the field in an accelerated reference frame. Moreover, SRE and entanglement can be harvested at the same time, including the particularly strong form of non-local SRE.

We have also shown that, at least with this  protocol,  it is not possible to harvest non-locality from the field, since there is either not enough entanglement or enough SRE. Even starting with a maximally entangled state, one can harvest SRE in non-local form, but a trade-off of resources happens so that the CHSH inequalities can never be violated. This depends on the fact that only non-local SRE is extracted, and in a symmetric way. This gives the main ingredient to formulate a no-go theorem for non-locality detection in a field, which we will address in future work.   This will require also a proof about the impossibility of violation of the Bell's inequalities for {\em mixed} states with maximal non-local magic.

In perspective, one wonders how steering and the Reeh-Schlieder theorem can be used to transfer state preparation in the field to state preparation of detectors\cite{RevModPhys.90.045003}.

It would also be interesting to study how the harvested SRE depends on the dimensionality of the detectors. Then one could also consider more realistic scenarios, for instance, studying the case of finite detectors rather than point like, or with different switching functions. Another interesting possibility is to consider massive scalar field moving along a trajectory in curved space-time instead of the simple setting of Minkowski background. 

\begin{acknowledgments}
The authors thank the anonymous referee whose comments have been invaluable in improving the scientific quality of this work. AH, JO and Stefano Cusumano acknowledge support from the PNRR MUR project PE0000023-NQSTI. AH acknowledges support from the PNRR MUR project CN 00000013-ICSC. Simone Cepollaro thanks Andrea Boccia and Michele Viscardi for useful discussions. 
We thank the authors of \cite{nystrom_2024_harvestingmagic} for valuable feedback.

\end{acknowledgments}
%

\newpage
\appendix
\onecolumngrid

\section{Computation of the reduced density matrix of the detectors\label{sec:app_den_mat}}

In this section, we show how to derive the reduced density matrix of the detectors when they are initialized in the state $\ket{00}$. For different choices of the initial state, the procedure goes exactly the same.

Let us introduce the following notation to shorten significantly the expressions:
\ba
\nonumber
\epsilon_{i}(t)=\epsilon(\tau_i(t))\frac{d\tau_i}{dt}, \quad \phi_i(t)=\phi(\bm{x}_i(t)),\quad {\phi_i}^{\pm}&=&\int dt \epsilon_i(t) e^{\pm i \Omega \tau_i(t)} {\phi}_i,\quad \ket{E_i^{\pm}}={\phi}_i^{\pm}\ket{0}_{\phi} 
\ea
where $i \in \{A,B\}$.
Using this notation, the interaction Hamiltonian reads:
\begin{equation}
    {H}_i=\lambda \epsilon_{i}(t)(e^{i \Omega \tau_{i}(t)}\ket{1}\bra{0}+e^{-i \Omega \tau_{i}(t) }\ket{0}\bra{1}) \otimes {\phi}_{i}(t) 
\end{equation}
The evolution occurs in both proper times $\tau_A$ and $\tau_B$, and for small value of the coupling constant $\lambda << 1$, the final state can be expanded as
\be
\ket{\psi_f}={U}\ket{\psi_i}=\sum_{n}\lambda^{n}\ket{\psi_{f}^{(n)}}
    \label{final state}
\ee 
To show this explicitly, we write $U$ in terms of the coordinate time $t$, with respect to which the vacuum state of the field is defined.

Before showing the perturbative terms of the final state $\ket{\psi_f}$, let us introduce the additional notation that will be useful in the following: 
\ba
\nonumber
    \phi_i^{\pm}(t)=\int_{-\infty}^{t} dt' \epsilon_i(t') e^{\pm i \Omega \tau_i(t')} {\phi}_i(t'),\qquad \ket{E_i^{\pm}(t)}=\phi_i^{\pm}(t) \ket{0}_{\phi} 
\ea
We can then compute the various terms of the series expansion.
\begin{itemize}
    \item $O(\lambda^0)$
   \ba
       \ket{\psi_{f}^{(0)}}&=&{\mathbbm{1}}\ket{\psi_{i}}=\ket{00}\ket{0}_{\phi}
   \ea
   \item $O(\lambda^1)$
\ba
    \ket{\psi_{f}^{(1)}}&=&-i \int dt \Bigl(\frac{d\tau_A}{dt} {H}_{\rm int}^{(A)} (\tau_A(t)) + \frac{d\tau_B}{dt} {H}_{\rm int}^{(B)} (\tau_B(t))\Bigr) \ket{\psi_{i}}\nonumber\\
    &=&-i \lambda \int dt \Bigl[\Bigr.\epsilon_{A}(t) (e^{i \Omega \tau_{A}(t)}\ket{1}\bra{0}+e^{-i \Omega \tau_{A}(t) }\ket{1}\bra{0}) \otimes {\phi}_{A}(t)\nonumber\\
    &+& \epsilon_{B}(t)(e^{i \Omega \tau_{B}(t)}\ket{0}\bra{1}+e^{-i \Omega \tau_{B}(t) }\ket{1}\bra{0}) \otimes {\phi}_{B}(t) \Bigl.\Bigr] \ket{00}\ket{0}_{\phi} \nonumber\\
   \nonumber &=& -i \lambda \int dt \Bigl[ \Bigr.\epsilon_{A}(t) e^{i \Omega \tau_{A}(t)}\ket{10} \otimes {\phi}_{A}(t) \ket{0}_{\phi} \nonumber\\
    &+&\epsilon_{B}(t)e^{i \Omega \tau_{B}(t)}\ket{01} \otimes {\phi}_{B}(t) \ket{0}_{\phi}\Bigl.\Bigr] \\
    &=&-i\lambda\left(\ket{10}\ket{E_A^+}+\ket{01}\ket{E_B^+}\right) 
    \ea
\item  $O(\lambda^2)$
\ba
\nonumber
\ket{\psi_f^{(2)}}&=&-\frac{\lambda^2}{2}\int dt\int dt' \mathcal{T}\Biggl[\frac{d\tau_A}{dt}\frac{d\tau_A}{dt'}H_{\rm int}^{(A)}(\tau_A(t))H_{\rm int}^{(A)}(\tau_A(t'))+\frac{d\tau_A}{dt}\frac{d\tau_B}{dt'}H_{\rm int}^{(A)}(\tau_A(t))H_{\rm int}^{(B)}(\tau_B(t')) \\
&+&\frac{d\tau_B}{dt}\frac{d\tau_A}{dt'}H_{\rm int}^{(B)}(\tau_B(t))H_{\rm int}^{(A)}(\tau_A(t'))+\frac{d\tau_B}{dt}\frac{d\tau_B}{dt'}H_{\rm int}^{(B)}(\tau_B(t))H_{\rm int}^{(B)}(\tau_B(t'))\Biggr]\ket{\psi_i}
\ea
Let us look first at one of these four integrals. For the term proportional to $H_A(\tau_A(t))H_A(\tau_A(t'))$ we get: 
\ba && \int dt \int dt' \frac{d\tau_A}{dt}\frac{d\tau_A}{dt'}H_{\rm int}^{(A)}(\tau_A(t))H_{\rm int}^{(A)}(\tau_A(t')) \ket{\psi_i}= \int dt \int_{-\infty}^t dt'\frac{d\tau_A}{dt}\frac{d\tau_A}{dt'}H_{\rm int}^{(A)}(\tau_A(t))H_{\rm int}^{(A)}(\tau_A(t')) \ket{00}\ket{0}_{\phi} \nonumber\\
&=&\int dt \int_{-\infty}^t dt' \epsilon_A(t)\epsilon_A(t')\Big(e^{i\Omega \tau_A(t)}\ket{1}\bra{0}_A+e^{-i\Omega \tau_A(t)}\ket{0}\bra{1}_A\Big)\Big(e^{i\Omega \tau_A(t')}\ket{1}\bra{0}_A\nonumber \\
&+&e^{-i\Omega \tau_A(t')}\ket{0}\bra{1}_A\Big)\ket{00}\phi_A(t)\phi_A(t')\ket{0}_{\phi} \nonumber\\
&=&\int dt \int_{-\infty}^t dt' \epsilon_A(t)\epsilon_A(t') e^{-i\Omega \tau_A(t)}e^{i\Omega \tau_A(t')} \phi_A(t)\phi_A(t')\ket{0}_{\phi}\ket{00}\nonumber\\
&=&\int dt \epsilon_A(t)e^{-i\Omega \tau_A(t)}\phi_A(t)\ket{E_A^+(t)}\ket{00} \nonumber\\
&=& \ket{00}\phi_A^-\ket{E_A^+(t)}
\ea
Notice that the operator $\phi_A^-$ act on the state $\ket{E^+_A(t)}$ performing an integration over the time variable, ensuring that the resulting state would not depend on $t$.

The other combinations of Hamiltonians can be computed in the same way as above. One  can show that the final state at the second order in $\lambda$ is given by:
\be \ket{\psi_f^{(2)}}=-\lambda^2\Big(\ket{00}\phi_A^-\ket{E_A^+(t)}+\ket{11}\phi_A^+\ket{E_B^+(t)}+\ket{11}\phi_B^+\ket{E_A^+(t)}+\ket{00}\phi_B^-\ket{E_B^+(t)}\Big) \ee 

\end{itemize}

Since we are interested in harvesting resources from the detector state, we have to trace out the degrees of freedom of the field ${\rho}_{AB}=Tr_{\phi} \ket{\psi_{f}}\bra{\psi_{f}}$.
Taking terms up to the second order in $\lambda$, one obtains the following density matrix:
\ba \label{rho00}
{\rho}_{AB}=\begin{pmatrix}
1-\lvert E_A \rvert^2-\lvert E_B \rvert^2 & 0 & 0& M\\
0& \lvert E_B \rvert^2 & \bra{E_B^+}\ket{E_A^-} &0 \\
0&\bra{E_A^+}\ket{E_B^-}&\lvert E_A \rvert^2&  0 \\
M^*&0&0&0 \label{densitymatrix}
\end{pmatrix}+\order{\lambda^4}
\ea
where $\absval{E_i}^2$ is the probability of having the $i$-th detector excited after the interaction, $M= -(\bra{E_B^-}\ket{E_A^-}+\bra{E_A^-}\ket{E_B^-})/2$ is the probability that the two detectors exchange a virtual particle and $\bra{E_B^+}\ket{E_A^-}=\bra{E_A^+}\ket{E_B^{-}}^*$ is the overlap between the excited states of the two detectors. 

\section{Final state of the detectors for other initial states\label{sec:other_states}}
In this section we report the final state of the detectors given the initial states used in Sec.~\ref{sec:accel}, namely $\rho_{AB}(0)=\dyad{\Phi^+}$ and $\rho_{AB}((0)=(\ket{0}\otimes\ket{T})(\bra{0}\otimes\bra{T})$.

When the two detectors are initialized in the state $\dyad{\Phi^+}$, the final state can be written as:
\ba
\nonumber
&&\mathcal{E}(\dyad{\Phi^+})=\\
\nonumber
&&\frac{1}{2}\begin{pmatrix}
1-|E_A|^2-|E_B|^2+M+M^*&0&0&1-|E_A|^2-|E_B|^2+2M\\
0&|E_A|^2+|E_B|^2+2M'&C+C'&0\\
0&C^*+C^{'*}&|E_A|^2+|E_B|^2+2M''&0\\
1-|E_A|^2-|E_B|^2+2M^*&0&0&1-|E_A|^2 -|E_B|^2+M+M^*
\end{pmatrix}
\ea
where
\ba
\nonumber
M'=\frac{\braket{E_A^-}{E_B^-}+\braket{E_B^+}{E_A^+}}{2}\\
\nonumber
M''=\frac{\braket{E_A^+}{E_B^+}+\braket{E_B^-}{E_A^-}}{2}\\
\nonumber
C=\braket{E_A^-}{E_B^+}+\braket{E_B^+}{E_A^-}\\
\nonumber
C'=\braket{E_A^-}{E_A^-}+\braket{E_B^+}{E_B^+}
\ea
Notice also that $M'+M''=-M-M^*$, so that the final state has trace 1. From this matrix elements one can also derive a perturbative expression fore the expectation value of the Bell operator, obtaining:
\ba
\nonumber
\Tr\left[\mathcal{B}_0\mathcal{E}(\dyad{\Phi^+})\right]=2-3(|E_A|^2+|E_B|^2)+6\Re[M]+\Re[C+C']
\ea

When the two detectors are initialized in the state $\ket{0}\otimes\ket{T}$, where $\ket{T}=\frac{\ket{0}+e^{-i\pi/4}\ket{1}}{\sqrt{2}}$, one obtains the final state with matrix elements:
\ba
\nonumber
&&\rho_{11}=\frac{1}{2}\left(1-|E_A|^2\right)\\
\nonumber
&&\rho_{12}=\frac{e^{i\pi/4}}{2}\left(1 - |E_A|^2-|E_B|^2-i\braket{E_B^-}{E_B^-}\right)\\
\nonumber
&&\rho_{13}=\frac{1}{2}\left(-e^{i\pi/4}\frac{C}{2}+e^{-i\pi/4}\braket{E_B^-}{E_A^-}\right)\\
\nonumber
&&\rho_{14}=\left(\frac{\braket{E_B^-}{E_A^-}-\braket{E_A^-}{E_B^-}}{4}\right)\\
\nonumber
&&\rho_{22}=\frac{1}{2}\left(1-|E_A|^2\right)\\
\nonumber
&&\rho_{23}=\frac{\braket{E_B^+}{E_A^-}-\braket{E_A^-}{E_B^+}}{4}\\
\nonumber
&&\rho_{24}=\frac{1}{2}\left(e^{i\pi/4}\braket{E_B^+}{E_A^-}+e^{-i\pi/4}M\right)\\
\nonumber
&&\rho_{33}=\frac{|E_A|^2}{2}\\
\nonumber
&&\rho_{34}=\frac{e^{i\pi/4}|E_A|^2}{2}\\
\nonumber
&&\rho_{44}=\frac{|E_A|^2}{2},
\ea
where the other off-diagonal elements are just the complex conjugate of the ones shown, and we have chosen to list the elements rather than show the entire matrix for typographic reasons.
\section{Three different acceleration scenarios\label{sec:app_accel}}
In this section we give the expression for the spacetime trajectories in the three different acceleration scenarios  of the two detectors, i.e., \textit{parallel} $a_{\parallel}$, \textit{antiparallel} $a_\nparallel$ and \textit{perpendicular} $a_\perp$, and show how the matrix elements of $\rho_{AB}$ only depend on the acceleration in all three cases.

\subsection{Parallel acceleration}
Suppose the two detectors are accelerated along the $x$-direction with acceleration $a_\parallel$ and initial mutual distance $L$ fixed to $1$. The trajectories of both detectors have the $y$ ad $z$ coordinates set to 0 along the motion, while the $t$ and $x$ coordinates are written in terms of the proper time as:
\ba 
t_A&=&\frac{\sinh a_\parallel\tau_A}{a_\parallel}, \\
x_A&=&\frac{L}{2}+\frac{\cosh a_\parallel\tau_A -1}{a_\parallel},\\
t_B&=&\frac{\sinh a_\parallel\tau_B}{a_\parallel}, \\
x_B&=&-\frac{L}{2}+\frac{\cosh a_\parallel\tau_B -1}{a_\parallel}.
\ea 

Let us look at the expression of the term $\absval{E_A}^2$
\ba 
\absval{E_A}^2 =  \bra{E_A^+}\ket{E_A^-} 
=  \int\int  e^{-\frac{\tau_A}{2 \sigma^2}} e^{-\frac{\tau'_A}{2\sigma^2}} e^{-i \Omega (\tau_A -\tau'_A)}W(\tau_A,\tau_A', a_{\parallel}) d\tau_A d\tau_A' \label{E_int}
\ea
where 
\ba W(\tau_A,\tau_A', a_{\parallel}) & := & \bra{0}\phi_A(x(\tau_A),t(\tau_A))\phi_A(x(\tau_A'), t(\tau_A'))\ket{0} \\
& = & -\frac{1}{4\pi^2}\frac{1}{(t(\tau_A)-t(\tau_A')-i\epsilon)^2-\absval{x(\tau_A)-x(\tau_A')}^2}.
\ea
is the Wightman function of the massless scalar field:

This function depends on the proper times and on the acceleration according to the trajectories of the detectors. Finally, after the integrations in \eqref{E_int}, the term $\absval{E_A}^2$ is only a function of the acceleration.

In a similar way, all the matrix elements can also be derived as integral functions depending on the acceleration parameter.
For example, the off-diagonal term:
\ba 
\langle E_A|E_B\rangle & = & \int \int e^{-\frac{\tau_A^2}{2\sigma^2}} e^{-\frac{\tau_B^2}{2 \sigma^2}} e^{i\Omega(\tau_A+\tau_B)} W(\tau_A,\tau_B,a_{\parallel}) d\tau_A d\tau_B
\ea

\subsection{Anti-parallel acceleration}
\begin{figure}[!t] 
\centering
\begin{subfigure}{0.33\textwidth}
\centering
\includegraphics[width=\linewidth]{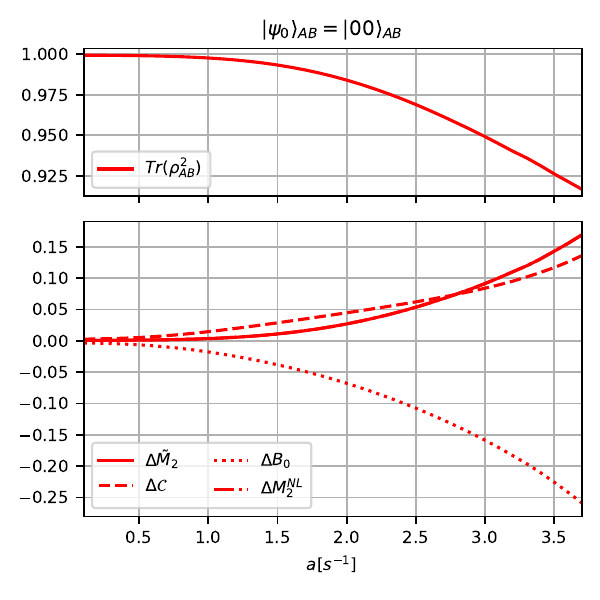}
\caption{}
\label{fig:00_antipar}
\end{subfigure}
\begin{subfigure}{0.33\textwidth}
\centering
\includegraphics[width=\linewidth]{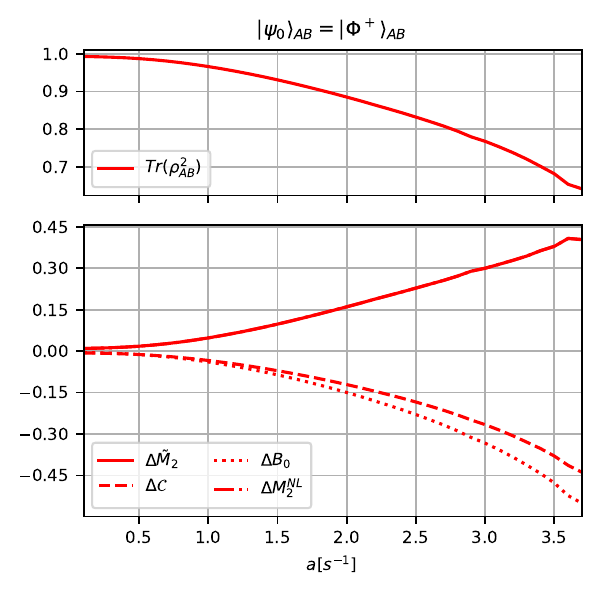}
\caption{}
\label{fig:bell_antipar}
\end{subfigure}
\begin{subfigure}{0.33\textwidth}
\centering
\includegraphics[width=\linewidth]{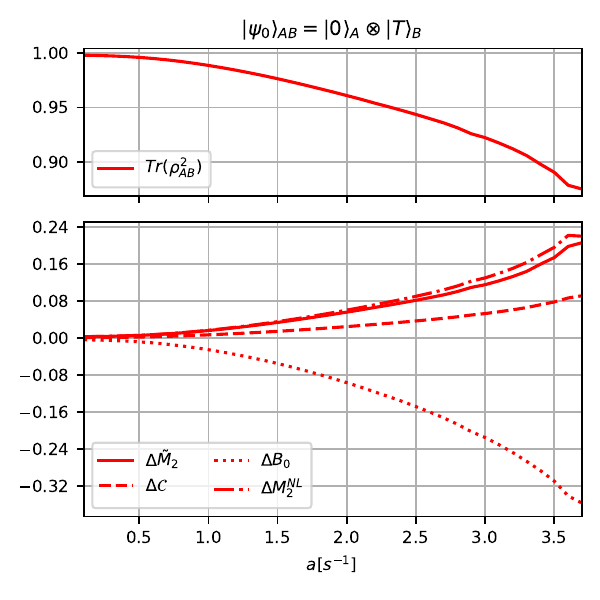}
\caption{}
\label{fig:0T_antipar}
\end{subfigure}
\caption{Plots of the harvested magic (local and non local) and entanglement, together with the expectation value of the Bell operator, as a function of the acceleration $a$ in the case of antiparallel acceleration $a_\nparallel$. Panel~\ref{fig:00_antipar} shows the evolution of the harvested resources when the detectors start in the state $\ket{\psi_0}_{AB}=\ket{00}_{AB}$. Panel~\ref{fig:bell_antipar} shows the harvested resources when the detectors start in the Bell state $\ket{\Phi^+}_{AB}$. Panel~\ref{fig:0T_antipar} shows the harvested resources when the detectors are initialized in the state $\ket{\psi_0}_{AB}=\ket{0}_A\otimes\ket{T}_B$. The upper plots show the evolution of the purity of the state of the detectors.}
\label{fig:plots_antip}
\end{figure}
The setting for anti-parallel acceleration $a_\nparallel$ is the same as the previous case, but one of the two detectors has the spatial coordinate $x$ with inverted sign:
\ba 
t_A&=&\frac{\sinh a_\nparallel\tau_A}{a_\nparallel}, \\
x_A&=&\frac{L}{2}+\frac{\cosh a_\nparallel\tau_A -1}{a_\nparallel},\\
t_B&=&\frac{\sinh a_\nparallel\tau_B}{a_\nparallel}, \\
x_B&=&-\frac{L}{2}-\frac{\cosh a_\nparallel\tau_B -1}{a_\nparallel}.
\ea 

With the same calculations shown for the parallel case, all the matrix elements of the detectors final state are written as integrals of the Wightman functions and after numerical integration over the proper times, all the terms will functions of the acceleration $a_{\nparallel}$. 
In \cref{fig:plots_antip}, we show the harvested magic and entanglement together with the expectation value of the Bell operator for detectors with antiparallel acceleration.

\subsection{Perpendicular acceleration}
\begin{figure}[!t]
\centering
\begin{subfigure}{0.33\textwidth}
\centering
\includegraphics[width=\linewidth]{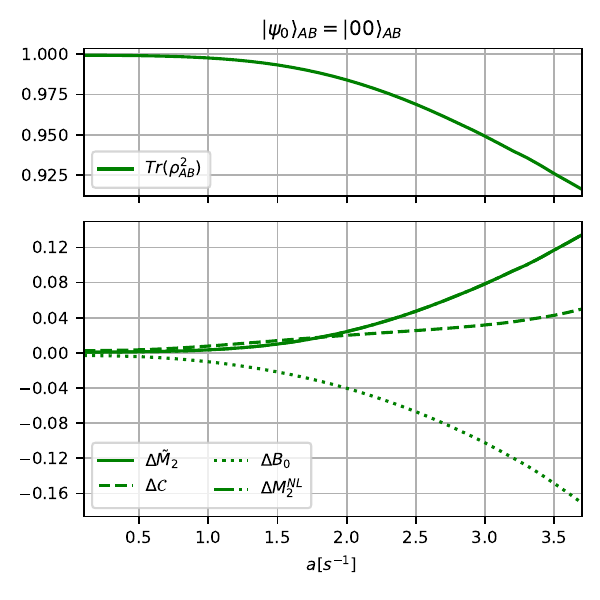}
\caption{}
\label{fig:00_perp}
\end{subfigure}
\begin{subfigure}{0.33\textwidth}
\centering
\includegraphics[width=\linewidth]{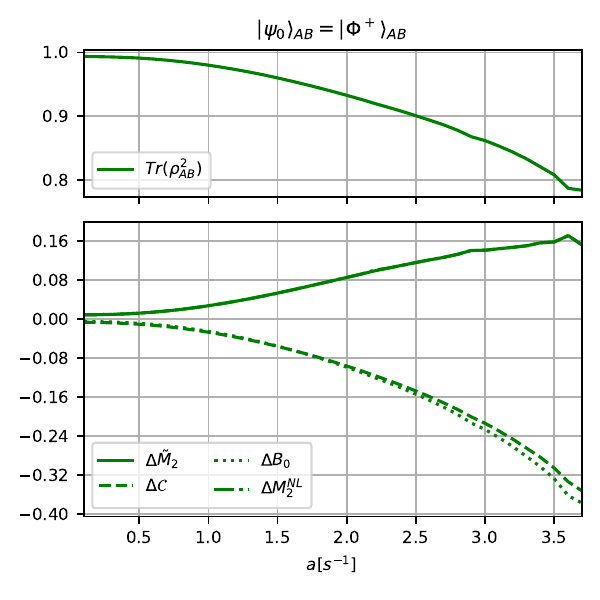}
\caption{}
\label{fig:bell_perp}
\end{subfigure}
\begin{subfigure}{0.33\textwidth}
\centering
\includegraphics[width=\linewidth]{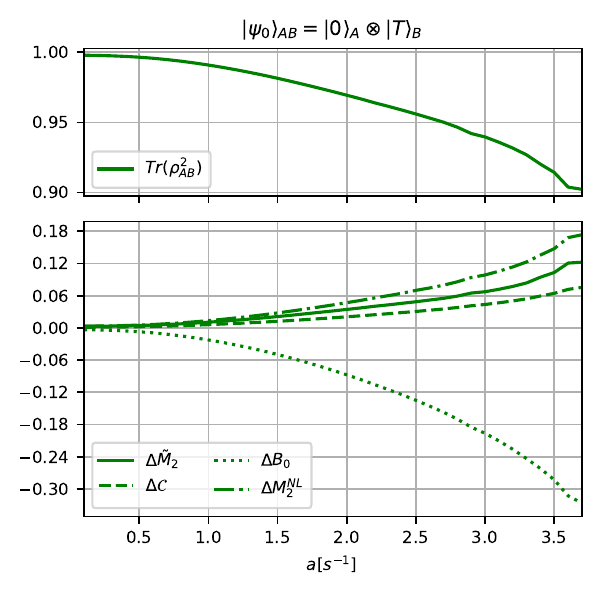}
\caption{}
\label{fig:0T_perp}
\end{subfigure}
\caption{Plots of the harvested magic (local and non local) and entanglement, together with the expectation value of the Bell operator, as a function of the acceleration $a_\perp$ in the case of perpendicular acceleration. Panel~\ref{fig:00_perp} shows the evolution of the harvested resources when the detectors start in the state $\ket{\psi_0}_{AB}=\ket{00}_{AB}$. Panel~\ref{fig:bell_perp} shows the harvested resources when the detectors start in the Bell state $\ket{\Phi^+}_{AB}$. Panel~\ref{fig:0T_perp} shows the harvested resources when the detectors are initialized in the state $\ket{\psi_0}_{AB}=\ket{0}_A\otimes\ket{T}_B$. The upper plots show the evolution of the purity of the state of the detectors.}
\label{fig:plots_perp}
\end{figure}

Without loss of generality, we can take spacetime trajectories of detectors in perpendicular acceleration $a_\perp$ as:
\ba 
t_A&=&\frac{\sinh a_\perp\tau_A}{a_\perp}, \\
y_A&=&\frac{\cosh a_\perp\tau_A -1}{a_\perp},\\
t_B&=&\frac{\sinh a_\perp\tau_B}{a_\perp}, \\
x_B&=&\frac{\cosh a_\perp\tau_B -1}{a_\perp}+L.
\ea

In \cref{fig:plots_perp}, we show the harvested magic and entanglement together with the expectation value of the Bell operator for detectors with perpendicular acceleration.
\end{document}